\documentclass[a4paper,fleqn]{cas-sc}
\usepackage[authoryear,longnamesfirst]{natbib}

\def\tsc#1{\csdef{#1}{\textsc{\lowercase{#1}}\xspace}}
\tsc{WGM}
\tsc{QE}
\begin{document}
\let\WriteBookmarks\relax
\def\floatpagepagefraction{1}
\def\textpagefraction{.001}

\shorttitle{}    

\shortauthors{M. Rahaman et~al.}  

\title [mode = title]{
%A higher-order compact difference method sheds new insights to miscible viscous fingering of a finite slice
HOC simulations of miscible viscous fingering of a finite slice: A new insight
}  




\author[1]{Mijanur Rahaman}%[<options>]

% \cormark[1]


% \ead{}


\credit{Formal analysis, Investigation, Methodology, Validation, Visualization, Writing - original draft}

\affiliation[1]{organization={Department of Mathematics, Indian Institute of Technology Guwahati},
            addressline={}, 
            city={Guwahati},
%          citysep={}, % Uncomment if no comma needed between city and postcode
            postcode={781039}, 
            state={Assam},
            country={India}}

\author[1]{Jiten Chandra Kalita}%[]

% Footnote of the second author

% Email id of the second author
% \ead{}


\credit{Formal analysis, Methodology, Visualization, Resources, Supervision, Writing - original draft, Writing - review \& editing}

% \affiliation[2]{organization={},
% %          citysep={}, % Uncomment if no comma needed between city and postcode

\author[1]{Satyajit Pramanik}%[]
\cormark[1]
% Footnote of the second author

% Email id of the second author
\ead{satyajitp@iitg.ac.in}


\credit{Conceptualisation, Formal analysis, Investigation, Methodology, Resources, Supervision, Validation, Visualization, Writing - original draft, Writing - review \& editing}

% \affiliation[2]{organization={},
% %          citysep={}, % Uncomment if no comma needed between city and postcode

\cortext[1]{Corresponding author}



\begin{abstract}
% Here goes the abstract \nocite{*}%% Remove this line from your manuscript.
We investigate the dynamics of viscous fingering (VF) in miscible slices in homogeneous, isotropic porous media. The fluid flow is governed by incompressible Darcy's law, whereas the solute transport is described using an advection-diffusion equation. The viscosity of the miscible system depends on the solute concentration, creating a viscosity contrast between the displacing fluid and the finite sample. When expressed in terms of stream function, the flow is described by a system of nonlinear, two-way coupled advection-diffusion type equations. We consider three types of boundary conditions: (a) periodic, (b) impermeable (zero normal velocity) and no-flux (solute), and (c) permeable (allowing non-zero normal velocity) and no diffusive flux (solute) transverse boundaries. This initial boundary value problem is solved numerically using a fourth-order compact finite difference method, while the Crank-Nicolson technique is used for time integration. Although the onset of viscous fingering and early time behavior are independent of the choice of boundary types, long-time behavior, solute mixing and spreading depend on the boundary conditions. In particular, it is observed that the permeable boundaries allow solute mass to increase, leading to stronger fingering instabilities, larger mixing lengths and non-trivial evolution of interfacial lengths. The findings of this study have implications in chromatography separation. 
\end{abstract}


%\item Higher-order compact difference method has been used to solve miscible viscous fingering instability in a finite slice. 
%\item The coupled nonlinear advection-diffusion-reaction type equations are solved subject to three different types of boundary conditions. The physical significance of each of these boundary types are explained. 
%\item It is shown that boundary conditions play crucial role in the dynamics of the solute. Implications in the context of real-life applications are highlighted. 

% Keywords

\maketitle

% Main text
\section{Introduction}\label{sec:intro}

Fluid flow and mixing in porous media are active areas of research devoted to characterizing industrial and environmental processes, such as oil recovery \citep{homsy1987viscous}, carbon dioxide sequestration \citep{huppert2014fluid}, and chromatography separation \citep{de2005viscous}. Fluid flow in porous media can feature hydrodynamic instabilities such as Saffman-Taylor \citep{saffman1958penetration} and Rayleigh-Taylor instabilities \citep{riaz2006onset}, depending on the flow configuration and physical properties of the flow. When a viscous fluid filling the voids in a porous medium is driven forward by the pressure of another driving fluid, the interface between them is likely to be unstable if the driving fluid is less viscous than the other. This instability is known as the Saffman-Taylor instability. As the occurrence of this instability leads to the formation of a finger-shaped structure of the upstream fluid invading the downstream one, it is also called viscous fingering (VF), which can be observed for both miscible and immiscible fluids \citep{saffman1958penetration}.

Viscous fingering is traditionally associated with petroleum engineering, but it has garnered increasing attention in environmental and chemical applications. Notably, VF plays a significant role in the spreading of contaminants in aquifers, mixing during brine transport, and in the band broadening observed in liquid chromatography, where a finite-width sample is displaced by a bulk fluid through a porous medium. In systems involving a single interface between two semi-infinite miscible fluids with a monotonic viscosity–concentration relationship, VF occurs when a less viscous fluid displaces a more viscous one, corresponding to a positive log-mobility ratio  $R>0$. However, for finite slices of fluid embedded within another, VF can arise at either the rear or front interface, depending on the sign of $R$ \citep{mishra2008differences}. The miscible viscous fingering (VF) problem has been investigated, governed by Darcy’s law coupled with a solute transport equation, where viscosity depends on concentration. Numerical studies frequently employ the Fourier pseudo-spectral method \citep{tan1988simulation, mishra2008differences,pramanik2016fingering} due to its exponential accuracy. However, this method inherently requires periodic boundary conditions, limiting the exploration of more physically realistic settings. There exist several well-established numerical techniques in the existing literature for dealing with the same. A higher-order discontinuous Galerkin (DG) method was used to simulate VF problems for miscible displacement in porous media \citep{li2016numerical, li2015high} and compared with cell-centred finite volume (CCFV) in terms of computational time and grid orientation effect. In another work, the streamfunction-vorticity approach has been adopted for the simulation, where the spectral Galerkin and a compact sixth-order finite difference scheme have been used for stream function and concentration, respectively \citep{hidalgo2012scaling}. Fully alternating-direction implicit (ADI) technique combined with a Hartley-based pseudo-spectral method has been used for numerical computation by \citet{islam2005fully}. 

Linear stability analysis, as well as the nonlinear dynamics of the miscible viscous fingering of a more viscous sample displaced by a low one, have been studied by several groups of researchers \citep{rousseaux2007viscous, kim2012linear, de2005viscous}. \citet{de2005viscous} found that fingering of finite slices is a transient phenomenon due to the decrease over time in the viscosity contrast across the interface induced by the fingering and dispersion process. They also analyzed the spreading characteristics in the context of contaminant spreading underground and the sample evolution using a chromatographic column. Subsequently, based on linear stability results, \citet{rousseaux2007viscous} identified the most probable growth rates of small perturbations and their corresponding wavelengths. Later, \citet{mishra2008differences} expanded this work to explore the influence on the extent of the sample for the case of a low viscous sample surrounded and displaced by a high viscous fluid. Their findings revealed that, due to the flow direction, the finger that developed at the frontal interface had a significant impact on the variance or mixing length. In contrast, for a more viscous sample, where VF formed at the rear interface, the impact on mixing was less pronounced. In their study, they implement the periodic boundary conditions.

While these assumptions are convenient, they may not fully capture the behavior observed in experimental setups or real-world applications. Recent works have explored the influence of more realistic or mixed boundary conditions, including Neumann-type fluxes and no-flow lateral boundaries, revealing nontrivial effects on interface dynamics and finger growth. In this direction \citet{kumar2019boundary} studied the effect of inlet boundary conditions by taking two different type of conditions -- (a) the absorbing boundary condition (Dirichlet type) in which a constant concentration is prescribed at the inlet, and (b) reflective boundary condition (Neumann type) $(\partial c/\partial x = 0)$ is used at inlet boundary. They are inspired by the experimental VF works, which are performed by filling the Hele-Shaw cell with more viscous fluid and then injecting other less viscous fluid through the inlet boundary. Their study reveals that the onset time of VF is earlier for the reflective case than the adsorbing one. They also found a threshold value of $R$, at which the onset is most delayed between these two cases. \citet{pramanik2015effect} studied the influence of P\'eclet number in a two-dimensional flow with no-flux boundary conditions at the lateral boundaries using linear stability analysis as well as nonlinear simulations. \citet{kim2023miscible} studied the effect of lateral boundary conditions in a cylindrical packed column on miscible VF and found that there is a strong dependence on the onset and growth of VF. Therefore, a systematic understanding of transverse boundary conditions is not reported. 

In this work, we are interested in understanding the viscous fingering phenomenon when a finite-width sample of fluid of different viscosity is being displaced by another lower-viscosity fluid \citep{mishra2008differences}. Here, we specifically address the effect of transverse boundary conditions other than the periodic, thus relaxing the assumption of periodicity to examine how different boundary conditions influence the development of viscous fingering and the other flow properties. By analyzing the resulting flow patterns and solute transport under various boundary configurations, we aim to understand the role of domain boundaries in shaping instability growth and mixing dynamics.

This work utilizes a temporally second-order and spatially fourth-order accurate HOC finite difference method \citep{kalita2002class, kalita2001hoc} on a uniform grid to numerically study the rectilinear displacement of a miscible high/low viscous sample. This reconstructed scheme enables simulations over a broader range of parameter values and allows for subsequent analysis. We utilize the streamfunction-vorticity formulation of the governing equations, which includes a convection-diffusion equation for solute concentration coupled with Darcy's law for fluid velocity, while adhering to the incompressibility condition of the fluid. 

The outline of this article is as follows. Mathematical formulation of the problem is presented in section \ref{sec:slice_formulation}, followed by numerical methods to solve the model equations. The results are discussed in section \ref{sec:slice_results} before summarizing the article in section \ref{sec:slice_conclusion}. 

\section{Mathematical formulation}  \label{sec:slice_formulation}

\begin{figure}
% \begin{center}
\centering
    \includegraphics[width = 0.8\textwidth]{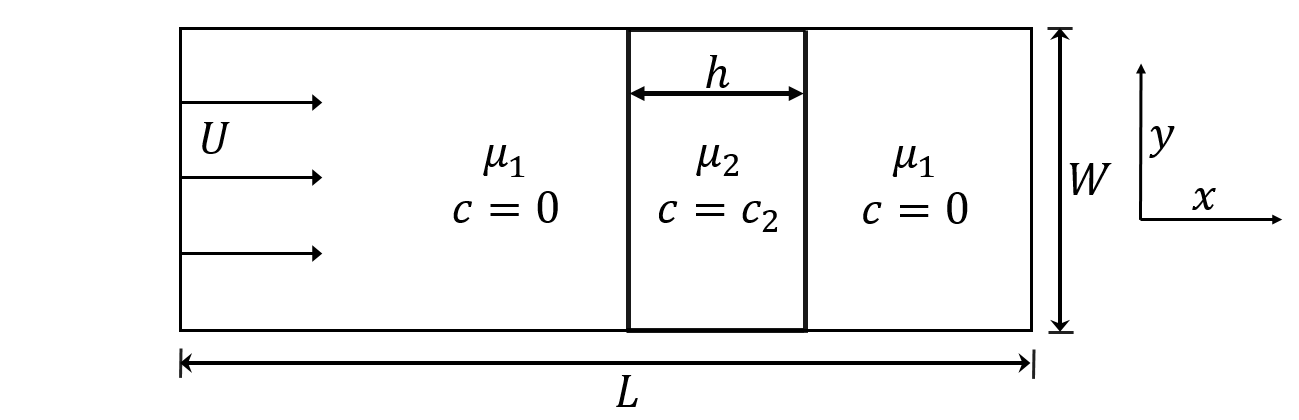}
    \caption{Schematic representation of the problem under consideration. A finite sample of viscosity $\mu_2$ is displaced by another fluid of viscosity $\mu_1$ in a homogeneous, isotropic porous medium (not to be scaled). The displacing fluid is injected at a velocity $(U, 0)$.} 
% \end{center}
\label{fig:slice_schematic}
\end{figure}

We consider rectilinear displacements of a miscible rectangular sample of size $h \times W$ in a two-dimensional, homogeneous, isotropic porous medium of dimension $L \times W$, with a constant permeability $\kappa$. The concentration of the solute is denoted by $c$, which follows a mass balance equation of advection-diffusion type. A viscous fluid of viscosity $\mu_1$ and solute concentration $c = 0$ is injected at a constant speed $U$ along the $x$-direction to displace the finite sample (see figure \ref{fig:slice_schematic} for a schematic). The viscosity of the finite sample is $\mu_2$, which depends on the solute concentration that is assumed to be $c = c_2$ initially. The fluids are assumed to be incompressible, neutrally buoyant, non-reactive, and contact miscible \citep{mishra2008differences, pramanik2015effect}. The flow and transport of the solute are governed by a system of two-way coupled, nonlinear partial differential equations as follows: 
\begin{eqnarray}
    \label{eq:darcy_law}
    & & \boldsymbol{\nabla}\cdot\boldsymbol{u} = 0, \qquad \boldsymbol{u} = -\frac{\kappa}{\mu(c)}\boldsymbol\nabla p, \\ 
    \label{eq:transport}
    & & \frac{\partial c}{\partial t} + \boldsymbol \nabla \cdot (\boldsymbol{u} c) = D \nabla^2 c, 
\end{eqnarray}
where $p$ is the hydrodynamic pressure, $\boldsymbol{u} = (u, v)$ is the two-dimensional velocity vector, $\mu$ is the dynamic viscosity of the fluid, and $D$ corresponds to the diffusion co-efficient of solute concentration in the ambient fluid. 

%%%%%%%%%%%%%%%%%%%%%%%%%%%%%%%%%%%%%%%%%%%%%%%%%%%%%%%%%%%%%%%%%%%%%%%
% figure- [schematic of the problem]
%%%%%%%%%%%%%%%%%%%%%%%%%%%%%%%%%%%%%%%%%%%%%%%%%%%%%%%%%%%%%%%%%%%%%%%%

% \subsection{Governing equations and non-dimensional formulation} \label{subsec:non-dimensional}

% We consider the porous medium with low porosity where the porescale Reynolds number is negligible and flow is governed by Darcy's law, 
% \begin{equation}
%     \label{eq:darcy_law}
%     \boldsymbol{\nabla}\cdot\boldsymbol{u} = 0, \qquad \boldsymbol{u} = -\frac{\kappa}{\mu(c)}\boldsymbol\nabla p, 
% \end{equation}
% where $p$ is the hydrodynamic pressure, $\boldsymbol{u} = (u, v)$ is the two-dimensional velocity vector, $\mu$ is the dynamic viscosity of the fluid. Assuming a homogeneous, isotropic dispersion of the solute in the solvent, transport of the solute is governed by an advection-diffusion equation
% \begin{equation}
%     \label{eq:transport}
%     \frac{\partial c}{\partial t} + \boldsymbol \nabla \cdot (\boldsymbol{u} c) = D \nabla^2 c, 
% \end{equation}
% where the isotropic dispersion coefficient $D$ is assumed to be constant. 

% To convert the equations \eqref{eq:darcy_law}-\eqref{eq:transport} into dimensionless form, the characteristic length, velocity, time, pressure, viscosity and concentration are chosen as $D/U$, $U$, $D/U^2$, $\mu_1D/\kappa$, $\mu_1$, and $c_2$ respectively. 
% Hence, the non-dimensional variables are 
Diffusion relaxes the concentration gradients between the displacing fluid and the defending sample. Thus, the viscosity contrast between the underlying fluids, responsible for fingering instability, relaxes. To investigate the effects of various flow parameters on fingering dynamics and the subsequent spreading of the sample, we are interested in studying the dynamics relative to the diffusive length and time scales. In particular, we render the following scaling 
\begin{equation}
    \label{eq:scaling}
    (\tilde{x}, \tilde{y}) = \frac{(x, y)}{D/U}, \;\; \boldsymbol{\tilde{u}} = \frac{\boldsymbol{u}}{U}, \;\; \tilde{t} = \frac{t}{D/U^2}, \;\; \tilde{p} = \frac{p}{\mu_1 D/\kappa}, \;\; \tilde{\mu} = \frac{\mu}{\mu_1}, \;\; \tilde{c} = \frac{c}{c_2}, 
\end{equation}
to non-dimensionlize equations \eqref{eq:darcy_law}--\eqref{eq:transport}, resulting (after dropping the tilde symbols for notational simplicity)
\begin{eqnarray}
\label{eq:nondim_Darcy}
& & \boldsymbol\nabla\cdot\boldsymbol{u} = 0,  \qquad \boldsymbol\nabla p = -\mu(c)(\boldsymbol{u} + \boldsymbol{i}), \\ 
\label{eq:nondim_transport}
& & \frac{\partial c}{\partial t} + \boldsymbol\nabla\cdot(\boldsymbol{u}c) = \nabla^2c, 
\end{eqnarray}
in a reference frame moving with the injection velocity.
% , are  
Here, $\boldsymbol{i}$ is the unit vector in the $x$-direction. Following \citep[see][and reference therein]{mishra2008differences}, we assume  
\begin{equation}
    \label{eq:arrhenius}
    \mu(c) = e^{Rc} \qquad \mbox{where} \qquad R = \ln\left( \frac{\mu_2}{\mu_1} \right).
\end{equation}
Thus, a more (less) viscous sample displaced by a less (more) viscous invading fluid is represented by $R > 0$ ($R < 0$). Equations \eqref{eq:nondim_Darcy}--\eqref{eq:arrhenius} should be supplemented with appropriate initial and boundary conditions to complete the mathematical description of the problem. 

Equations \eqref{eq:nondim_Darcy}--\eqref{eq:nondim_transport} can be expressed in the following stream function formulation, 
\begin{eqnarray}
    \label{eq:concentration}
  & & \frac{\partial c}{\partial t} + \frac{\partial \psi}{\partial y}\frac{\partial c}{\partial x} - \frac{\partial \psi}{\partial x}\frac{\partial c}{\partial y} = \nabla^2c,  \\ 
    \label{eq:streamfunction}
   & &  \nabla^2\psi = -R\boldsymbol \nabla c \cdot{(\boldsymbol{\nabla}\psi + \boldsymbol{j})}, 
\end{eqnarray}
where $\psi(x,y,t)$ is the stream function such that $\displaystyle \boldsymbol{u}=(u,v)=\left(\frac{\partial \psi}{\partial y},-\frac{\partial \psi}{\partial x}\right)$, $\boldsymbol{j}$ is the unit vector in $y$-direction.
The dimensionless length and width of the porous medium become $L_x = UL/D$ and $L_y = UW/D$, whereas the sample has dimensionless length $l = Uh/D$. For completeness of the formulation, it is essential to state the initial and boundary conditions, which are discussed in \S \ref{subsec:initial_conditions} and \S \ref{subsec:boundary_conditions}, respectively.

\subsection{Initial condition} \label{subsec:initial_conditions}

In order to observe the long time behaviors of the finger pattern, the finite sample has been placed in $\displaystyle \left( \frac{4L_x}{5}- \frac{l}{2}, \frac{4L_x}{5} + \frac{l}{2} \right)$ for $R>0$, whereas for $R<0$ initial position of the sample is taken as $\displaystyle \left( \frac{L_x}{5} - \frac{l}{2}, \frac{L_x}{5} + \frac{l}{2} \right)$.
Therefore, the initial condition of concentration for the case $R>0$, can be written as, 
\begin{equation}
\label{eq:IC1}
	c(x,y,0)=
	\begin{cases}
		0, \qquad  \displaystyle x<\frac{4L_x}{5}-\frac{l}{2}, \\[0.8 em]
		1, \qquad  \displaystyle x\in \left( \frac{4L_x}{5}-\frac{l}{2}, \frac{4L_x}{5}+\frac{l}{2} \right),\\[0.8 em]
		0, \qquad  \displaystyle x>\frac{4L_x}{5}+\frac{l}{2},
	\end{cases}
\end{equation}
and for $R<0$ is,
\begin{equation}
\label{eq:IC2}
	c(x,y,0)=
	\begin{cases}
		0, \qquad   \displaystyle x<\frac{L_x}{5}-\frac{l}{2}, \\[0.8 em]
		1, \qquad    \displaystyle x\in \left( \frac{L_x}{5}-\frac{l}{2}, \frac{L_x}{5}+\frac{l}{2} \right),\\[0.8 em]
		0, \qquad  \displaystyle x>\frac{L_x}{5}+\frac{l}{2},
	\end{cases}
\end{equation}
whereas for the streamfunction, it is given by
\begin{equation}
\label{eq:ICpsi}
\psi(x,y,0)=0.
\end{equation}
To initiate the instability in the system, a random perturbation is added at both the rear and frontal interfaces. The noise added at the fronts: \(c = \left(  1\pm Ar \right)/2 \), where $r$ is a random number drawn from uniform distribution in $\left[0, 1\right]$ and $A = 10^{-3}$ is the noise amplitude. This noise triggers the fingering instability in a controlled manner and the hydrodynamic instability remains unaffected by the numerical discretization. A larger noise amplitude $A$ (e.g., $10^{-1}$) accelerates the onset of instability, whereas a smaller one delays the same. To compare patterns developing in the co-flow and counter-flow directions, the leading front should be initialized with $(1-c)$ if the rear is initialized with $c$, ensuring symmetric initial perturbations.

\subsection{Boundary conditions} \label{subsec:boundary_conditions}
As mentioned earlier, in this work, we investigate the effects of different boundary conditions on the fingering patterns and both qualitative and quantitative dynamics of the solute transport. For that purpose, the same initial random perturbation has been maintained in all the cases. The three different types of boundary conditions used in this study are detailed below:
\begin{eqnarray}
	& & \left.\boldsymbol{\nabla}{c}\cdot\boldsymbol{n}\right|_{x= 0,\, L_x}=0,  \left.\boldsymbol{\nabla}{ \psi}\cdot\boldsymbol{n}\right|_{x= 0,\, L_x}=0 \mbox{ and } \left.c\right|_{(x,0,t)}=\left.c\right|_{(x,L_y,t)}, \left.\psi\right|_{(x,0,t)}=\left.\psi\right|_{(x,L_y,t)}.
	\label{bc:Type-I} \\ 
	& & \left.\boldsymbol{\nabla}{c}\cdot\boldsymbol{n}\right|_{x= 0, \, L_x}=0, \left.\boldsymbol{\nabla}{ \psi}\cdot\boldsymbol{n}\right|_{x= 0,\, L_x}=0 \mbox{ and } \left.\boldsymbol{\nabla}{c}\cdot\boldsymbol{n}\right|_{y= 0,\, L_y}=0, \left.\psi\right|_{y= 0,\, L_y}=0.
	\label{bc:Type-II} \\
    & & \left.\boldsymbol{\nabla}{c}\cdot\boldsymbol{n}\right|_{x= 0,\, L_x}=0, \left.\psi\right|_{x= 0, \,L_x}=0 \mbox{ and } \left.\boldsymbol{\nabla}{c}\cdot\boldsymbol{n}\right|_{y=0,\, L_y}=0,  \left.\boldsymbol{\nabla}{ \psi}\cdot\boldsymbol{n}\right|_{y=0,\, L_y}=0.
	\label{bc:Type-III}
\end{eqnarray}
Boundary conditions \eqref{bc:Type-I} correspond to the periodic boundary conditions at the transverse boundaries. These are the most widely used boundary conditions in the literature \cite{de2005viscous, mishra2008differences, pramanik2015effect}. Boundary conditions \eqref{bc:Type-II} represent impermeable boundaries with no solute flux, whereas boundary conditions \eqref{bc:Type-III} correspond to permeable boundaries allowing normal flow without diffusive solute flux. 
For convenience, we designate these three boundary conditions mentioned in equations \eqref{bc:Type-I}, \eqref{bc:Type-II}, and \eqref{bc:Type-III} as Type-I, Type-II, and Type-III, respectively, throughout the article. 

\section{Numerical method}\label{sec:numerical}
    
A closed-form analytical solution is not available for the nonlinear system of two-way coupled advection-diffusion equations \eqref{eq:concentration}-\eqref{eq:streamfunction}. We seek numerical solutions of these equations utilising a higher-order compact difference scheme \citep{rahaman2025towards}. In the following, we briefly describe the algorithm used to solve the governing equations. 
\begin{figure}
% \begin{center}
\centering
    \includegraphics[width = 0.95\textwidth]{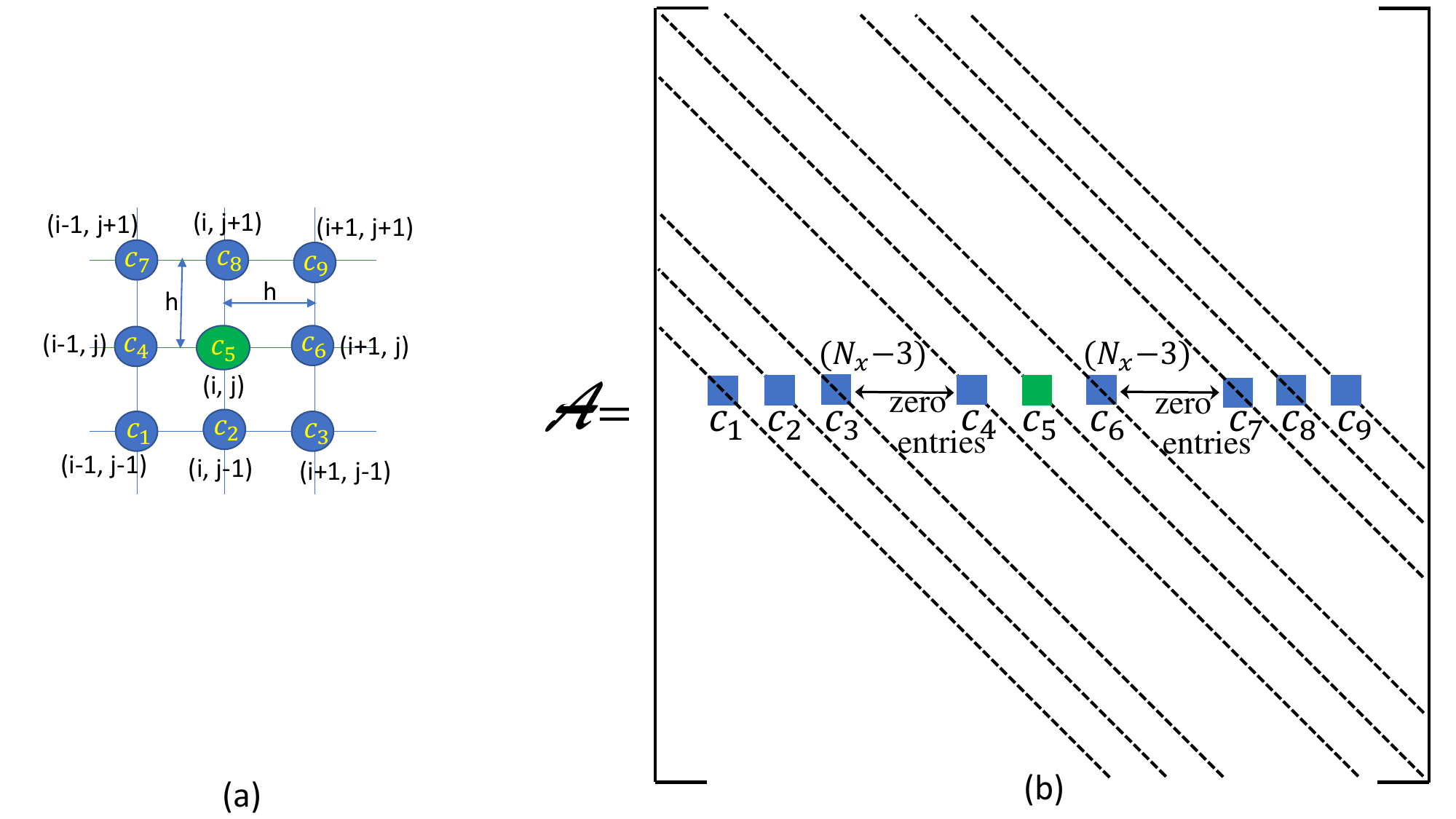}
    \caption{(a) The $9$-point HOC stencil with the associated coefficients in the matrix equation and (b) The structure of the coefficient matrix along with the locations of the non-zero coefficients.} 
% \end{center}
\label{fig:stencil}
\end{figure}

The spatial derivatives are discretized using fourth-order accurate finite difference approximations, while time integration is performed using a Crank-Nicolson technique, ensuring a second-order temporal accuracy. The computational domain $\displaystyle \left[ 0, L_x \right] \times \left[ 0, L_y \right] \times \left[ 0, T \right]$ is discretized using a uniform mesh of size $N_x \times N_y \times N_t$, where $N_x$, $N_y$ respectively are the number of grid points along $x$-, $y$-directions of the spatial domain assumed rectangular and $N_t$ is the number of grid points along the temporal $t$-direction. We denote a generic dependent variable $\phi$ at $(x_i, y_j, t_n)$ as $\phi_{i,j}^{(n)}$, where $x_i = ih, \; y_j = jh$ and $t_n = n\Delta t$ for $i = 0, 1, \hdots, N_x-1$, $j = 0, 1, \hdots, N_y-1$ and $n = 0, 1, \hdots, N_t-1$ with $h = L_x/(N_x-1) = L_y/(N_y-1)$ and $\Delta t = T/(N_t-1)$. Consequently, the HOC approximation of equation \eqref{eq:concentration} at the interior nodes ($i = 1, 2, \hdots, N_x - 2$, $j = 1, 2, \hdots, N_y - 2$) reads
\begin{eqnarray}
\label{eq:HOC_generalized} 
& & \left[ 1 + \frac{h^{2}}{12} \left( \delta_{x}^{2}+\delta_{y}^{2}-b_{i j} \delta_{x}-d_{i j} \delta_{y} \right) \right] \left(c_{ij}^{(n+1)}-c_{ij}^{(n)} \right) \nonumber \\ 
& & \quad =\frac{\Delta t}{2} \left[ \alpha_{i j} \delta_{x}^{2} + \beta_{i j} \delta_{y}^{2}-B_{i j} \delta_{x}-D_{i j} \delta_{y} + \frac{h^{2}}{6} \left( \delta_{x}^{2} \delta_{y}^{2}-b_{i j} \delta_{x} \delta_{y}^{2}-d_{i j} \delta_{x}^{2} \delta_{y}-\gamma_{i j} \delta_{x} \delta_{y} \right) \right] \left( c_{ij}^{(n+1)}+c_{ij}^{(n)} \right) \nonumber \\ 
& & \qquad \qquad \qquad \qquad \qquad \qquad \qquad \qquad \qquad \qquad + \mathcal{O} (\Delta t^2, h^4),
\end{eqnarray}
where $\delta_{x},\delta_{y},\delta_{x}^{2},\delta_{y}^{2}$ are the first and second order central difference operators, 
% $h$ is uniform step length along $x$ and $y$ directions respectively, and $\Delta t$ is the uniform time step length. 
The coefficients  $b_{ij},d_{ij},B_{ij},D_{ij},\alpha_{i j},\beta_{i j}$ and $\gamma_{i j}$  are defined as follows:
	\begin{eqnarray}
		b_{ij}&=&\bigg(\frac{\partial \psi}{\partial y}\bigg)_{ij}^{(n)},\\
		d_{ij}&=&-\bigg(\frac{\partial \psi}{\partial x}\bigg)_{ij}^{(n)},\\
		B_{i j}&=&\bigg[1+\frac{h^{2}}{12}\left(\delta_{x}^{2}+\delta_{y}^{2}-b_{i j} \delta_{x}-d_{i j} \delta_{y}\right) \bigg]b_{i j}, \\
		D_{i j}&=&\bigg[1+\frac{h^{2}}{12}\left(\delta_{x}^{2}+\delta_{y}^{2}-b_{i j} \delta_{x}-d_{i j} \delta_{y}\right) \bigg]d_{i j},\\
	%	F_{i j}&=&\bigg[1+\frac{h^{2}}{12}\left(\delta_{x}^{2}+\delta_{y}^{2}-b_{i j} \delta_{x}-d_{i j} \delta_{y}\right) \bigg]f_{i j},\\
		\alpha_{i j}&=&1+\frac{h^{2}}{12}\left(b_{i j}^{2}-2 \delta_{x} b_{i j}\right), \\
		\beta_{i j}&=&1+\frac{h^{2}}{12}\left(d_{i j}^{2}-2 \delta_{y} d_{i j}\right)\\
		\gamma_{i j}&=&\delta_{x} d_{i j} +\delta_{y} b_{i j}-b_{i j} d_{i j}.
	\end{eqnarray}
Note that the compact fourth-order approximation of the spatial derivatives leads to a nine-point stencil, as shown in figure \ref{fig:stencil}(a). In figure \ref{fig:stencil}(b), we display the corresponding nonzero entries of the coefficient matrix arising from the discretization of the governing equations, together with the locations of the associated stencil points. Thus, equation \eqref{eq:HOC_generalized} at the $(N_x-2) \times (N_y-2)$ interior points can be written in the matrix form as
	\begin{equation}
		\mathcal{A_I}\boldsymbol{c}^{(n+1)}=\boldsymbol{f}(\boldsymbol{c_I}^{(n)}),
		\label{eq:matrix_interior}
	\end{equation}
where $\mathcal{A_I}$ is a rectangular matrix of dimension $(N_x-2)  (N_y-2)\times N_x N_y$, each row containing at most 9 non-zero elements, $\boldsymbol{c}^{(n+1)}$ is the unknown concentration vector at the $(n+1)^{\rm th}$ time level with $N_x N_y$ components and the right hand side is a vector of length $(N_x-2)  (N_y-2)$. 

Likewise, the matrix equation corresponding to the $2(N_x+N_y-2)$ boundary points (see also section \ref{subsec:periodic_boundary_implement}) can be written as 
\begin{equation}
		\mathcal{A_B}\boldsymbol{c}^{(n+1)}=\boldsymbol{f}(\boldsymbol{c_B}^{(n)}),
		\label{eq:matrix_boundary}
	\end{equation}
where  $\mathcal{A_B}$ is a rectangular matrix of dimension $2(N_x+N_y-2) \times N_x N_y$ and   the right hand side is a vector of length $2(N_x+N_y-2)$. Upon assembling the contributions from all interior and boundary points, the global system over the entire computational domain assumes the form
     \begin{equation}
		\mathcal{A}\boldsymbol{c}^{(n+1)}=\boldsymbol{f}(\boldsymbol{c}^{(n)}),
		\label{eq:matrix_system1}
	\end{equation}
where the coefficient matrix $\mathcal{A}$ is a square matrix of order $N_x N_y$ and the right hand side is a vector of length $N_x N_y$. For Dirichlet boundary conditions, or when Neumann boundary conditions are discretized using first-order approximations, the matrix $\mathcal{A}$ has the banded structure depicted in figure \ref{fig:stencil}(b) with one principal diagonal and eight sub-diagonals corresponding to the nine-point stencil. The three clusters of diagonals, ordered from left to right in figure\ref{fig:stencil}(b), are associated with the $(j-1)$, $j$ and  $(j+1)$ levels of the stencil in figure \ref{fig:stencil}(a), respectively.

After computing the concentration values at the current time level, the streamfunction $\psi$ is determined from the equation \eqref{eq:streamfunction} through the steady-state form of the schemes \citep[see][and references therein for further details]{kalita2002class, kalita2001hoc, rahaman2025towards}. Consequently, the HOC approximation to equation \eqref{eq:streamfunction} yields
\begin{eqnarray}
\label{eq:HOC_generalized2}
& & -\tilde{\alpha}_{i j}\delta_{x}^{2} \psi_{i j} - \tilde{\beta}_{i j}\delta_{y}^{2} \psi_{i j} + \tilde{B}_{i j} \delta_{x} \psi_{i j} + \tilde{D}_{i j} \delta_{y} \psi_{i j} - \frac{h^{2}}{6}\left[\delta_{x}^{2} \delta_{y}^{2}-\tilde{b}_{i j} \delta_{x} \delta_{y}^{2}-\tilde{d}_{i j} \delta_{x}^{2} \delta_{y}-\tilde{\gamma}_{i j} \delta_{x} \delta_{y}\right] \psi_{i j} \nonumber \\ 
& & \qquad \qquad \qquad \qquad \qquad \qquad \qquad \qquad \qquad \qquad \qquad \qquad =\tilde{F}_{i j} + \mathcal{O}(h^4),
\end{eqnarray}

where the coefficients  $\tilde{b}_{ij}, \; \tilde{d}_{ij}, \; \tilde{B}_{ij}, \; \tilde{D}_{ij}, \; \tilde{\alpha}_{i j}, \; \tilde{\beta}_{i j}$, $\tilde{\gamma}_{i j}$ and $\tilde{F}_{i j}$ are defined as 
\begin{eqnarray}
\label{eq:bij_tilde}
& & \tilde{b}_{ij} = -R\delta_{x}c_{ij}^{(n+1)}, \\
\label{eq:dij_tilde}
& & \tilde{d}_{ij} = -R\delta_{y}c_{ij}^{(n+1)}, \\
\label{eq:Bij_tilde}
& & \tilde{B}_{ij} = \left[ 1 + \frac{h^{2}}{12} \left( \delta_{x}^{2} + \delta_{y}^{2} - \tilde{b}_{i j} \delta_{x} - \tilde{d}_{i j} \delta_{y} \right)  \right] \tilde{b}_{i j}, \\
\label{eq:Dij_tilde}
& & \tilde{D}_{i j} = \left[ 1 + \frac{h^{2}}{12} \left( \delta_{x}^{2} + \delta_{y}^{2} - \tilde{b}_{i j} \delta_{x} - \tilde{d}_{i j} \delta_{y} \right) \right] \tilde{d}_{i j}, \\
\label{eq:Fij_tilde}
& & \tilde{F}_{i j} = R \left[ 1 + \frac{h^{2}}{12} \left( \delta_{x}^{2} + \delta_{y}^{2} - \tilde{b}_{i j} \delta_{x} - \tilde{d}_{i j} \delta_{y} \right) \right] \delta_{y}c_{i j}^{(n+1)}, \\ 
\label{eq:alphaij_tilde}
& & \tilde\alpha_{i j} = 1 + \frac{h^{2}}{12} \left( \tilde{b}_{i j}^{2} - 2 \delta_{x} \tilde{b}_{i j} \right), \\
\label{eq:betaij_tilde}
& & \tilde{\beta_{i j}} = 1 + \frac{h^{2}}{12} \left( \tilde{d}_{i j}^{2} - 2 \delta_{y} \tilde{d}_{i j} \right), \\ 
\label{eq:gammaij_tilde}
& & \tilde{\gamma_{i j}} = \delta_{x} \tilde{d}_{i j} + \delta_{y} \tilde{b}_{i j} - \tilde{b}_{i j} \tilde{d}_{i j}. 
\end{eqnarray} 

After assimilating equation \eqref{eq:HOC_generalized2} with the boundary approximations, the matrix form of the discretized governing equations for $\psi$ reduces to
\begin{equation}
    \label{eq:matrix_system2}
    \tilde{\mathcal{A}} \boldsymbol{\psi}^{(n+1)} = \tilde{\boldsymbol{f}}(\boldsymbol{c}^{(n+1)}), 
\end{equation}
where the coefficient matrix $\tilde{\mathcal{A}}$ is an square matrix of order $N_x N_y$, and $\boldsymbol{\psi}^{(n+1)}$ is the unknown stream function vector at $(n+1)^{\rm th}$ time level having $N_x N_y$ components. It is easy to see that the structure of $\mathcal{\tilde{A}}$ similar to $\mathcal{A}$. Once $\psi$ is known, the gradients of the streamfunction appearing in \eqref{eq:concentration} are computed using the standard central difference approximation.

Next, we extend the higher-order compact difference approximations at the boundary points. 
% this discussion to the treatment of the boundary points and how the structure of the row is going to be different under different boundary conditions in the coefficient matrix (e.g., $\mathcal{A})$. 
% Numerical treatments of the three different types of boundary conditions employed in this study are unique as discusssed below. 
% separate numerical treatment. 
To demonstrate their implementation, we chose a representative point at the bottom boundary, namely, $(x_i,y_0)$.
% that covers all types of boundary conditions used in this work. 
For a Dirichlet-type boundary condition, where the unknown variable $\phi$ is prescribed (e.g., $\phi = \Phi_b(x,0)$) at the transverse boundary, it can be included in the matrix form using 
\begin{eqnarray}
	\label{eq:Dirichlet(x,L_y)}
	\phi_{i,0}^{(n+1)} = (\Phi_b)_{i,0} \text{ for } i=1, 2, \hdots, N_x-2. 
\end{eqnarray}
% This obtained relation \eqref{eq:Dirichlet(x,L_y)} Thus, one row of the revised coefficient matrix contains only one non-zero entry on the diagonal, which is $1$, corresponding to the unknown $\phi_{i,0}$ in the matrix system $\mathcal{A}\boldsymbol{\phi}^{(n+1)}=\boldsymbol{b}$. 
For the Neumann type boundary condition, the normal derivative of the unknown variable $\phi$ vanishes, i.e., 
\begin{equation}
 \boldsymbol{\nabla }\phi\cdot\boldsymbol{n} = 0. 
\end{equation}
At the bottom boundary, say at $(x_i,y_0)$, this condition reduces to
\begin{eqnarray}
\left. \frac{\partial \phi}{\partial y}\right|_{(x_i,\, y_0)} = 0 \Rightarrow \phi_{i,0}^{(n+1)} - \phi_{i,1}^{(n+1)} =  0.
\end{eqnarray}
% The above finite difference approximation of Neumann boundary condition yields a row in the coefficient matrix with only two non-zero entries which are $+1$ and $-1$, corresponding to the unknown variable $\phi_{i,0}^{(n+1)}$ and $\phi_{i,1}^{(n+1)}$ in the matrix system $\mathcal{A}\boldsymbol{\phi}^{(n+1)} = \boldsymbol{b}$. 
In contrast, the periodic boundary condition has been implemented using the original differential equation itself. In the following, we provide a detailed account of the HOC discretization when a periodic boundary condition is used. 

\subsection{Implementation of periodic boundary conditions at transverse boundary} \label{subsec:periodic_boundary_implement}
\begin{figure}
% \begin{center}
\centering
    \includegraphics[width = 1.00\textwidth]{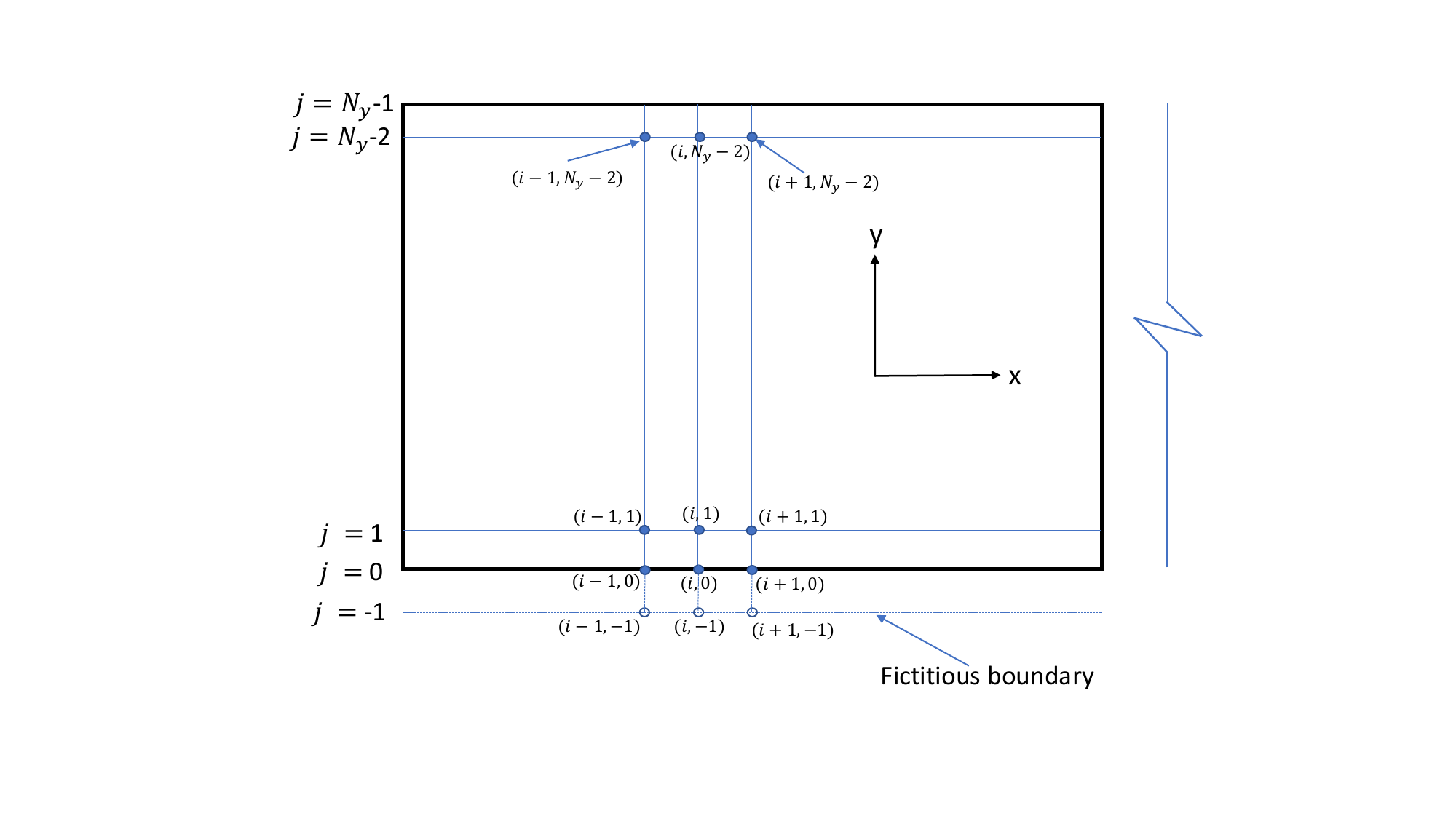}
    \caption{Illustration of the approximation of periodic boundary condition.} 
% \end{center}
\label{fig:stencilB}
\end{figure}
The periodic boundary conditions of a variable $\phi$ 
% in the transverse direction that has been employed under the Type-I condition. Periodicity of $\phi$, 
in a spatial rectangular domain $[0,L_x]\times[0,L_y]$ in the transverse direction can be mathematically expressed as, 
\begin{eqnarray}
	\phi(x, 0, t) = \phi(x, L_y, t), \qquad \forall x,\,\, \forall t \geq 0. 
	\label{eq:perioic_y}
\end{eqnarray}
The compact fourth-order $(9,9)$ stencil cannot be applied directly at periodic boundaries, as it introduces fictitious points near the corners and boundaries, shown as hollow circles in figure \ref{fig:stencilB}. Solid circles denote the actual computational stencil points. To resolve this, we employ the periodic boundary conditions to express the values at these fictitious points in terms of interior points. 

For the point $(x_i,y_0)$ (labeled as $(i, 0)$ in figure \ref{fig:stencilB}), on the bottom boundary $y = y_0$, the relation is 
\begin{eqnarray}
	\label{eq:discretization_periodic_(x_i,0)}
	& & a_1\phi_{i-1,N_y-2}^{(n+1)} + a_2\phi_{i,N_y-2}^{(n+1)} + a_3\phi_{i+1,N_y-2}^{(n+1)} + a_4\phi_{i-1,0}^{(n+1)} + a_5\phi_{i,0}^{(n+1)} + a_6\phi_{i+1,0}^{(n+1)} \nonumber\\
	& & \qquad\qquad\qquad\qquad\qquad\qquad + a_7\phi_{i-1,1}^{(n+1)} + a_8\phi_{i,1}^{(n+1)} + a_9\phi_{i+1,1}^{(n+1)} = g_{i,0}(\Phi^{(n)}), \;\; i = 1,2,..., N_x-2,
\end{eqnarray}
which is derived utilizing the periodicity relation $\phi_{k,-1}^{(n+1)} = \phi_{k,N_y-2}^{(n+1)}$, where $k=i-1,\;i,\;i+1$. As can be seen from equation \eqref{eq:discretization_periodic_(x_i,0)} and figure \ref{fig:stencilB}, the computational stencil at this boundary differs from that used at interior points (see figure \ref{fig:stencil}(a)). Consequently, in the actual matrix assembly for the iterative solvers, the coefficients $c_1,\;c_2\;{\rm and}\;c_3$ associated with figure \ref{fig:stencil}(b) are no longer applicable, and the corresponding coefficients at the level $j=N_y-2$ must be redefined. As a result, a typical row of the coefficient matrix associated with a point on this boundary contains three nonzero entries clustered around the principal diagonal, another three entries following $N_x-3$ zeros, and the final three entries located after $(N_x-1)(N_y-3)-2$ zeros. The coefficient matrix of the global system largely preserves the structure shown in figure \ref{fig:stencil}(b), except at those locations where the coefficients forming the clusters of diagonals on the left and right vanish. Moreover, due to conditions such as \eqref{eq:discretization_periodic_(x_i,0)}, additional small clusters of three diagonals may appear at certain positions beyond the nine diagonals depicted in figure \ref{fig:stencil}(b).

% This relation is used for all the points in the bottom boundary except the two end points, i.e., the relation is valid for $i = 1,2,..., N_x-2$. 

While we use the governing equations for the bottom boundary, for the top boundary, relation \eqref{eq:perioic_y} is utilized yielding
\begin{eqnarray}
	\label{eq:discretization_periodic_(x_i,L_y)}
	\phi_{i,N_y-1}^{(n+1)}-\phi_{i,0}^{(n+1)}&=&0, \text{ for } i=1,2,...,N_x-2
\end{eqnarray}
leading to a row in the coefficient matrix with only two non-zero entries, namely,  $+1$ and $-1$, corresponding to the unknowns at the points $(x_i, y_{N_y-1})$ and $(x_i, y_0)$.
 
Although we have not applied dual periodic boundary conditions to the dependent variable $\phi$, we would like to mention that for dual periodicity, the concept can be generalized to include periodicity of $\phi$ in the longitudinal direction in a similar fashion. Our discussion so far has been focused at a representative point of the transverse boundary; the same implementation strategy applies for all types of boundary conditions along the longitudinal boundaries.

\subsection{Solution of the system of linear equations}\label{subsec:linear_system}

Time marching of the coupled system is achieved by incorporating an inner and outer iterative procedure. The inner iterations are composed of solving the matrix equations \eqref{eq:matrix_system1} and \eqref{eq:matrix_system2} using the biconjugate gradient stabilized (BiCGStab) iterative solver. An incomplete LU decomposition is used as a preconditioner with the help of the Lis library \citep{lisnew}. The iterations are stopped when the norm of the residual vectors arising out of the respective system of equations falls below the tolerance limit $10^{-7}$. Computations of $c$, and $\psi$ employing equations \eqref{eq:matrix_system1}, and \eqref{eq:matrix_system2} constitute one outer iteration. The same is repeated at each time level until the desired final time is reached. The whole computational procedure can be summarized in the following algorithm.
\vspace{.3cm}
\hrule 
{\bf Algorithm} Computation of concentration and streamfunction 
\hrule
\begin{enumerate}
    \item {\bf Initialization:} Set parameters $R$ and $l$. Initialize  $n=0,\;t=0$ and then $\boldsymbol{c}^{(n)}=\boldsymbol{c}^{(0)}$, $\boldsymbol{\psi}^{(n)}=\boldsymbol{\psi}^{(0)}$ using \eqref{eq:IC1}-\eqref{eq:ICpsi}. 
    \item {\bf while} ($t<{\rm endTime}$) {\bf do}
    \begin{enumerate}
        \item Compute the gradients $\displaystyle \left (\frac{\partial \boldsymbol{\psi}}{\partial x}\right )^{(n+1)}$ and $\displaystyle \left (\frac{\partial \boldsymbol{\psi}}{\partial y}\right )^{(n+1)}$ using standard central difference approximations.
     
        \item Solve \eqref{eq:matrix_system1} by BiConjugate gradient stabilized method to compute $\boldsymbol{c}^{(n+1)}$ corresponding to \eqref{eq:concentration}.
        
        \item Substitute $\boldsymbol{c}^{(n+1)}$ obtained by {\bf step (b)} above and solve 
    \eqref{eq:matrix_system2} by BiConjugate gradient stabilized method to compute   $\boldsymbol{\psi}^{(n+1)}$ corresponding to \eqref{eq:streamfunction}.

         \item Set $\boldsymbol{c}^{(n)}=\boldsymbol{c}^{(n+1)}\; {\rm and}\;\boldsymbol{\psi}^{(n)}=\boldsymbol{\psi}^{(n+1)}$.
         \item Set $n=n+1,\; {\rm and} \;t=t+\Delta t$.
        
    \end{enumerate}
      {\bf end while}\\
      {\bf Output:} ${c}^{(n+1)}_{ij}=c(x_i,y_j,t_k)$, ${\psi}^{(n+1)}_{ij}=\psi(x_i,y_j,t_k)$ with $x_i=ih,\;y_j=jh\; {\rm and}\;t_k=k\Delta t$ for $i=0, 1,2,....,N_x-1$, $j=0,1,2,....,N_y-1$ and $k=1,2,....,N_t-1$.
\end{enumerate}
\hrule

\subsection{Grid independence and validation} \label{subsec:validation}

\begin{figure}
	\centering
	(a) \hspace{3.2 in} (b) \\ 
	\includegraphics[width=0.495\textwidth]{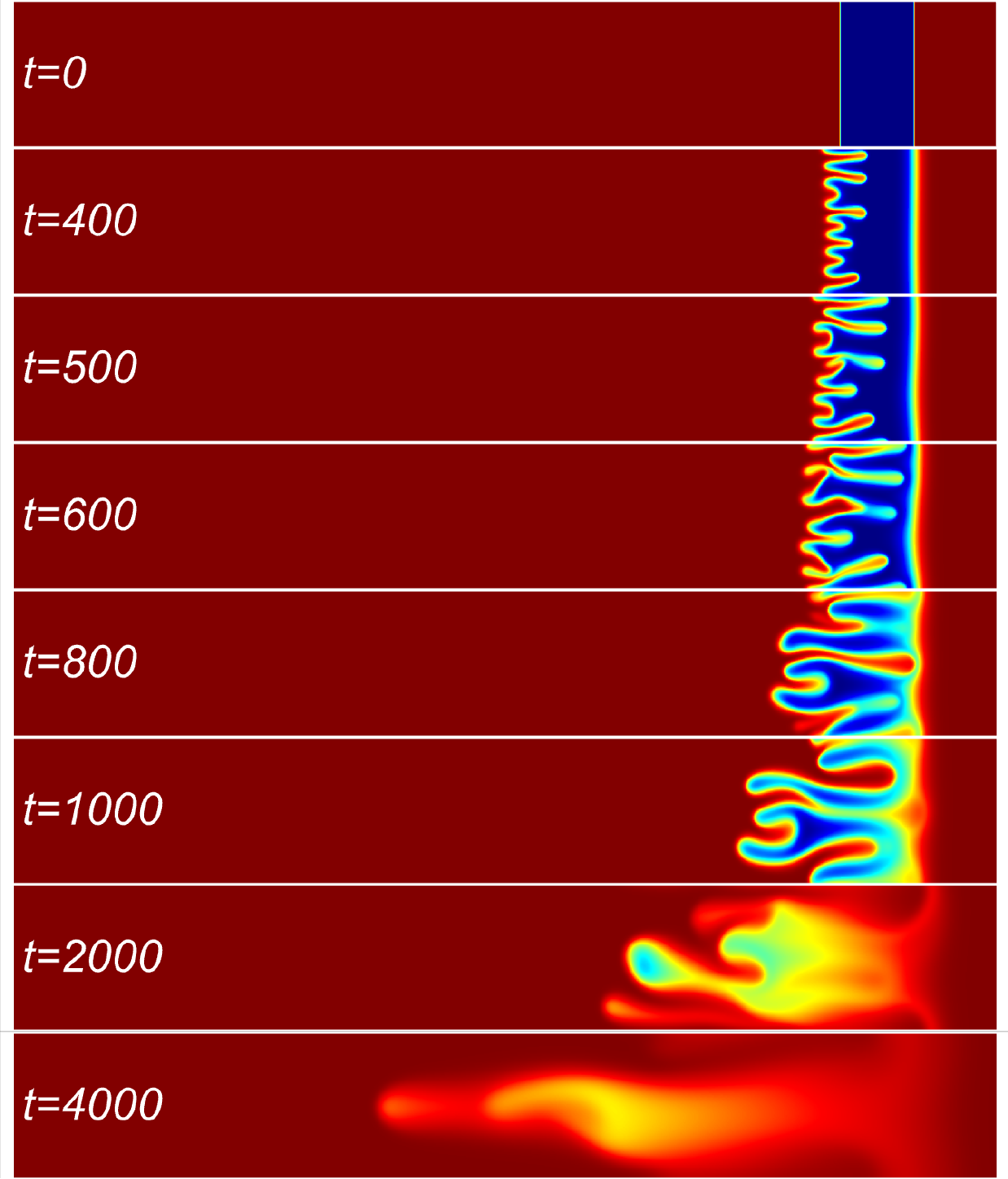} 
	\includegraphics[width=0.495\textwidth]{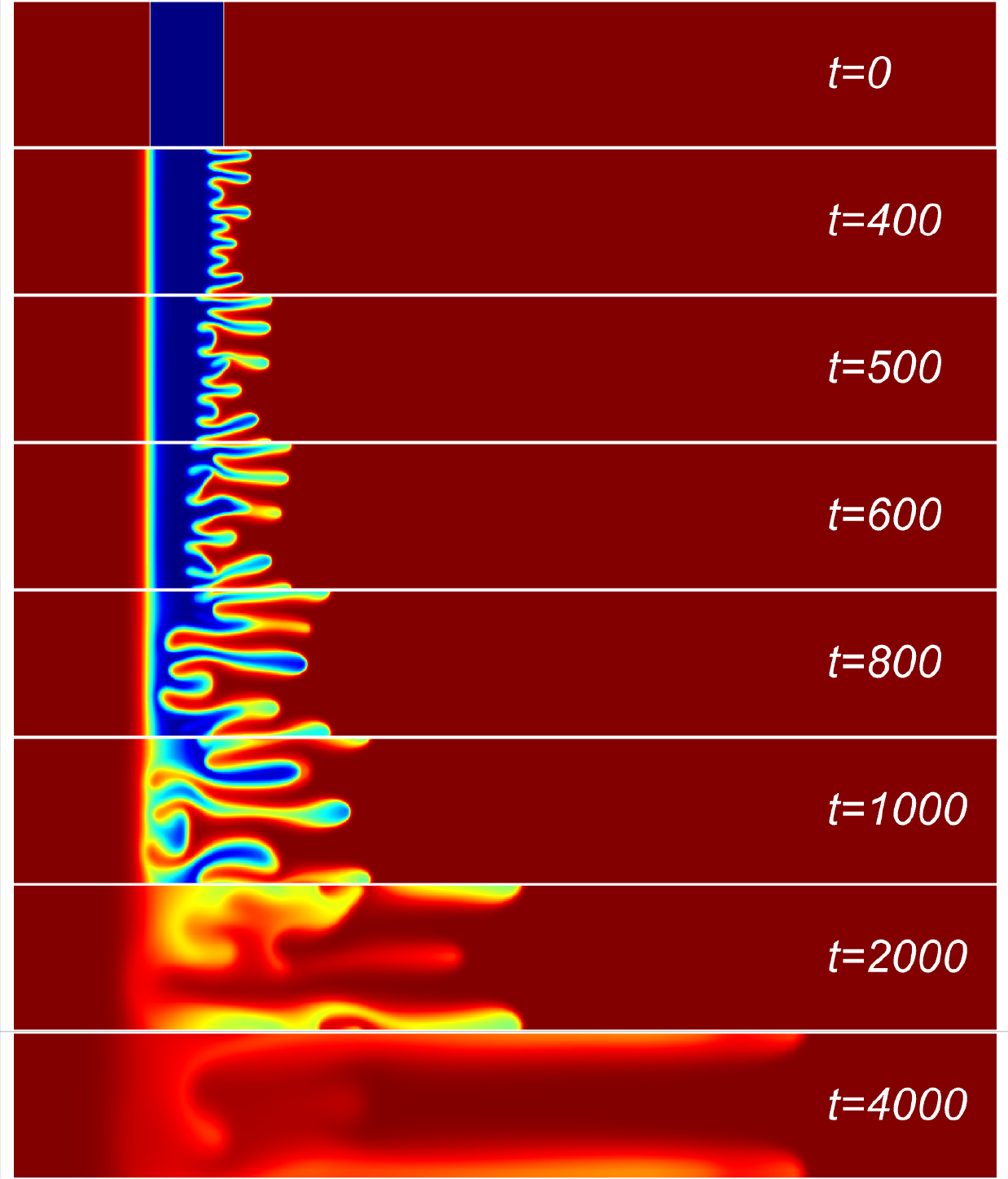}
	  \\[0.2cm]
	\includegraphics[width=0.5\textwidth]{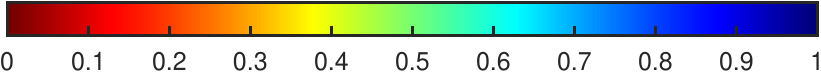}
	\caption{Spatio-temporal distribution of the finite sample of width $l = 256$ exhibiting fingering dynamics and solute spreading at various time levels for (a) $R = 3,$ (b) $R = -3$ corresponding to the Type-I boundary conditions. Other parameter are $L_x = 4096$, $L_y = 512$.} 
	\label{Fig:Type-I}
\end{figure}

In this section, we present grid independence and validation of the HOC method discussed above before proceeding to investigate fingering dynamics under different flow conditions. For all three types of boundary conditions, we choose $\lvert R \rvert  = 3$, sample width $l = 256$, $L_x = 4096$, $L_y = 512$. Three different spatial grid sizes, $513 \times 65$, $1025 \times 129$, and $2049 \times 257$, corresponding to $h = 8, 4$, and $2$, respectively, have been used along with a temporal step size $\Delta t = 0.2$. It is observed that increasing the grid size beyond $1025 \times 129$ does not alter the pattern formation. Therefore, all the computations shown in this work are performed using the spatial grid size $1025 \times 129$, i.e., $h = 4$ and temporal step size $\Delta t = 0.2$. For $R = 3$ and $R = -3$ with $L_y = 512$ and Type-I boundary conditions, the spatio-temporal evolution of the sample depicted in figure \ref{Fig:Type-I} exhibits an excellent agreement with the literature (see figure $2$ of \citet{mishra2008differences}). 

%%%%%%%%%%%%%%%%%%%%%%%%%%%%%%%%%%%%%%%%%%%%%%%%%%%%%%%%%%%%%%%%%%%%%%%%
% figure- [grid independence/validation] Type-I boundary conditions for signed R
%%%%%%%%%%%%%%%%%%%%%%%%%%%%%%%%%%%%%%%%%%%%%%%%%%%%%%%%%%%%%%%%%%%%%%%%

\section{Results and discussion} \label{sec:slice_results}

Here, we systematically discuss the fingering dynamics and spreading of a finite sample when the viscosity of the sample is more than ($R > 0$), equal to ($R = 0$), and less than ($R < 0$) that of the displacing fluid. 

\subsection{Transport of a finite sample in the absence of hydrodynamic instability} \label{sec:slice_viscosity_matched}

Here, we examine the case $R=0,$ which implies that there is no viscosity difference between the displacing and displaced fluids. Therefore, the finite sample and the ambient fluid move at the same velocity. Thus, in the moving frame of reference, fluid velocity is equal to zero, and a diffusion process governs the solute transport (see equation \eqref{eq:nondim_transport}). Therefore, in this case, the dynamics of the solute are independent of the boundary types considered here. To better understand the dynamics of miscible fluids, we resort to various qualitative and quantitative measures described below. The transverse-averaged concentration profile, a classical metric that has been widely utilized in experimental \citep{kretz2003experimental, bacri1991three} as well as in theoretical studies of VF \citep{mishra2008differences, pramanik2016fingering}, is defined as 
\begin{equation}
    \label{eq:trans_avg}
    \bar{c}(x,t)=\frac{1}{L_y}\int\limits_{0}^{L_y}c(x,y,t) {\rm d} y. 
\end{equation}
The first moment, also known as the centre of mass of the distribution, and its variance are defined as \citep{mishra2008differences}
\begin{equation}
    \label{eq:first_moment_x}
    m_x(t) = \frac{\int_{0}^{L_x}x\bar{c}(x,t) {\rm d} x}{\int_{0}^{L_x}\bar{c}(x,t) {\rm d} x},
\end{equation}
and
\begin{equation}
  \label{eq:var_x}
    \sigma_x^2(t) = \frac{\int_{0}^{L_x} x^2 \bar{c}(x,t) {\rm d} x}{\int_{0}^{L_x} \bar{c}(x,t) {\rm d} x} - \left[ \frac{\int_{0}^{L_x} x \bar{c}(x,t) {\rm d} x}{\int_{0}^{L_x} \bar{c} (x, t) {\rm d} x} \right]^2, 
\end{equation}
respectively. These measures help quantify the effects of diffusion in the longitudinal direction. 

%%%%%%%%%%%%%%%%%%%%%%%%%%%%%%%%%%%%%%%%%%%%%%%%%%%%%%%%%%%%%%%%%%%%%%%%
% figure- [(R = 0)]
%%%%%%%%%%%%%%%%%%%%%%%%%%%%%%%%%%%%%%%%%%%%%%%%%%%%%%%%%%%%%%%%%%%%%%%%
\begin{figure}
	\centering
	(a) \\ 
	\includegraphics[width=0.5\textwidth,height=6.1cm]{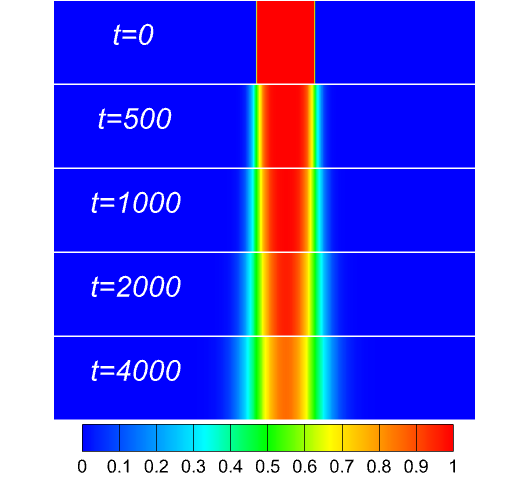} \\ 
	(b) \hspace{3.1 in} (c) \\
	\includegraphics[width=0.495\textwidth,height=6.6cm]{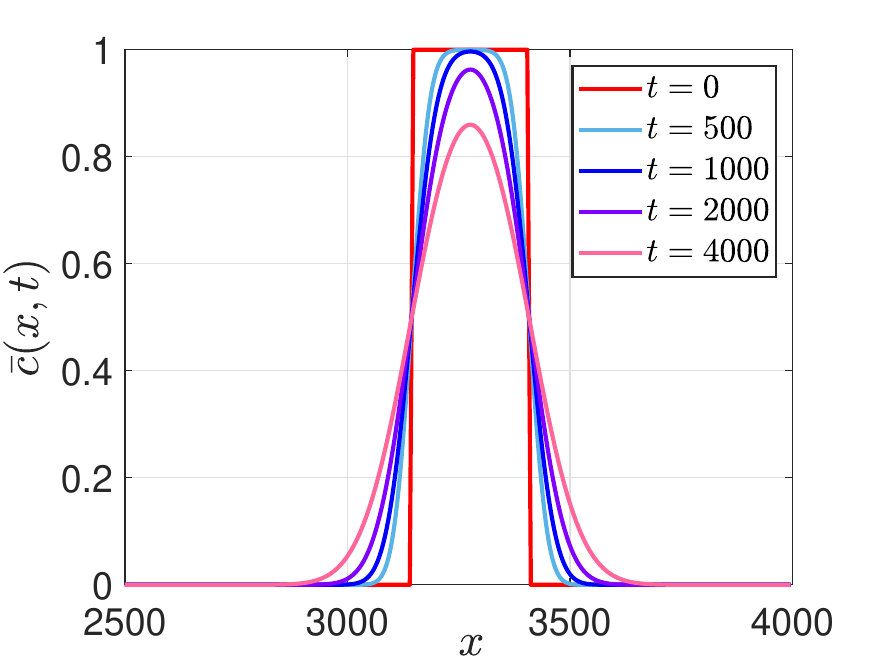} 
	\includegraphics[width=0.495\textwidth]{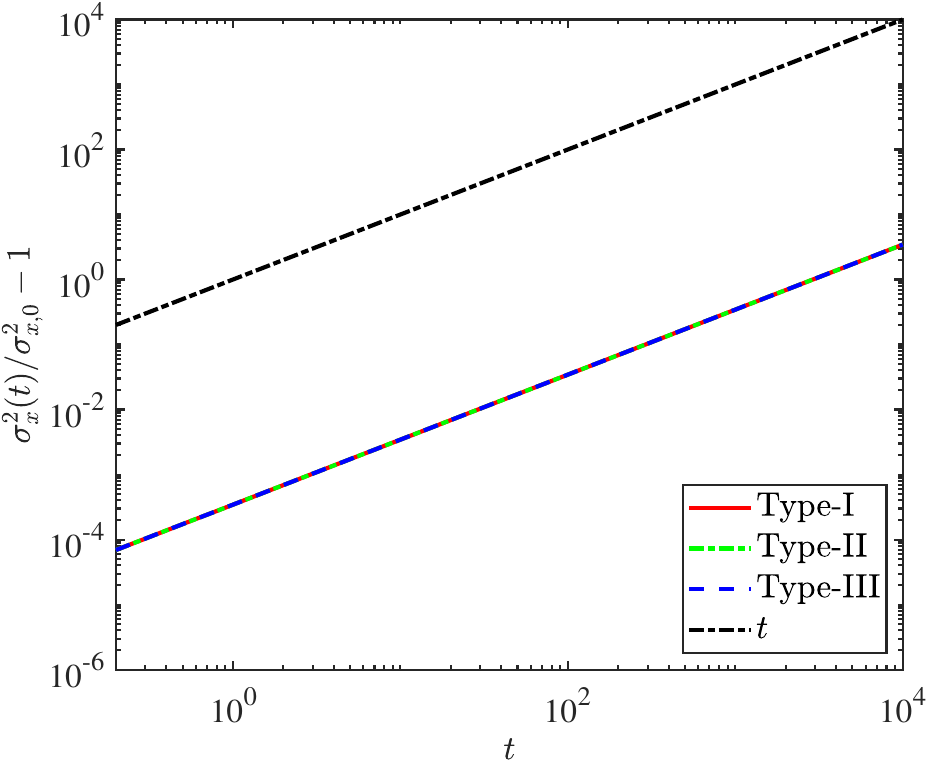} 
	\caption{(a) Spatio-temporal distribution of concentration for $R = 0$ at different time levels corresponding to the Type-I boundary conditions, (b) the corresponding transverse-averaged concentration. (c) Rescaled longitudinal variances, $(\sigma_x^2(t)/\sigma_{x, 0}^2 - 1)$, corresponding to the Type-I (red), the Type-II (green), and the Type-III (blue) boundary conditions are shown. It is observed that in the absence of viscous fingering, the dynamics are independent of the choice of the boundary conditions. In all three cases, the variances grow linearly in time. A black dash-dotted line is shown for reference.} 
	\label{Fig:R_0}
\end{figure} 

As anticipated, the solute sample diffuses isotropically and mixes with the displacing fluid. Due to the symmetry of the diffusive mixing about its mean position, the problem can be described in terms of one-dimensional diffusion, characterized by the variance that evolves linearly in time. Figure \ref{Fig:R_0}(a-b) demonstrates the diffusive spreading of the sample and the corresponding Gaussian profile of transverse-averaged concentration under Type-I boundary conditions. Figure \ref{Fig:R_0}(c) confirms that the longitudinal variance of the solute varies linearly with time, $\sigma_x^2(t)/\sigma_{x, 0}^2 - 1 \propto t$, and is indistinguishable between different boundary conditions considered in this study. 

\subsection{Viscous fingering of a finite slice}\label{subsec:VF_slice} 

%%%%%%%%%%%%%%%%%%%%%%%%%%%%%%%%%%%%%%%%%%%%%%%%%%%%%%%%%%%%%%%%%%%%%%%%
% R=3
%%%%%%%%%%%%%%%%%%%%%%%%%%%%%%%%%%%%%%%%%%%%%%%%%%%%%%%%%%%%%%%%%%%%%%%%
\begin{figure}
    \centering
    (a) \hspace{1.9 in} (b) \hspace{1.9 in} (c) \\ 
    \includegraphics[width=0.32\textwidth]{figures/fig4a.pdf}
    \includegraphics[width=0.32\textwidth]{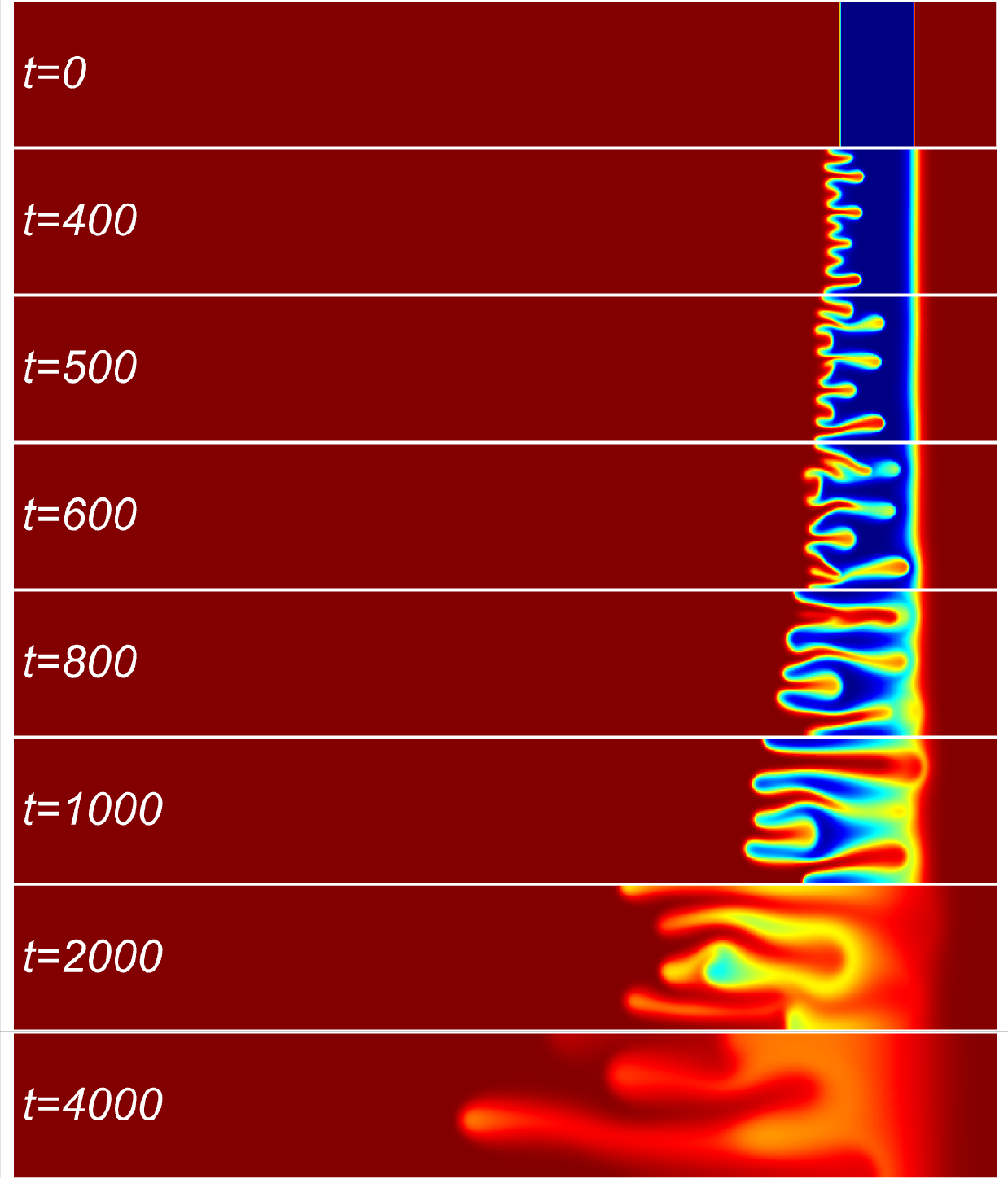}
    \includegraphics[width=0.32\textwidth]{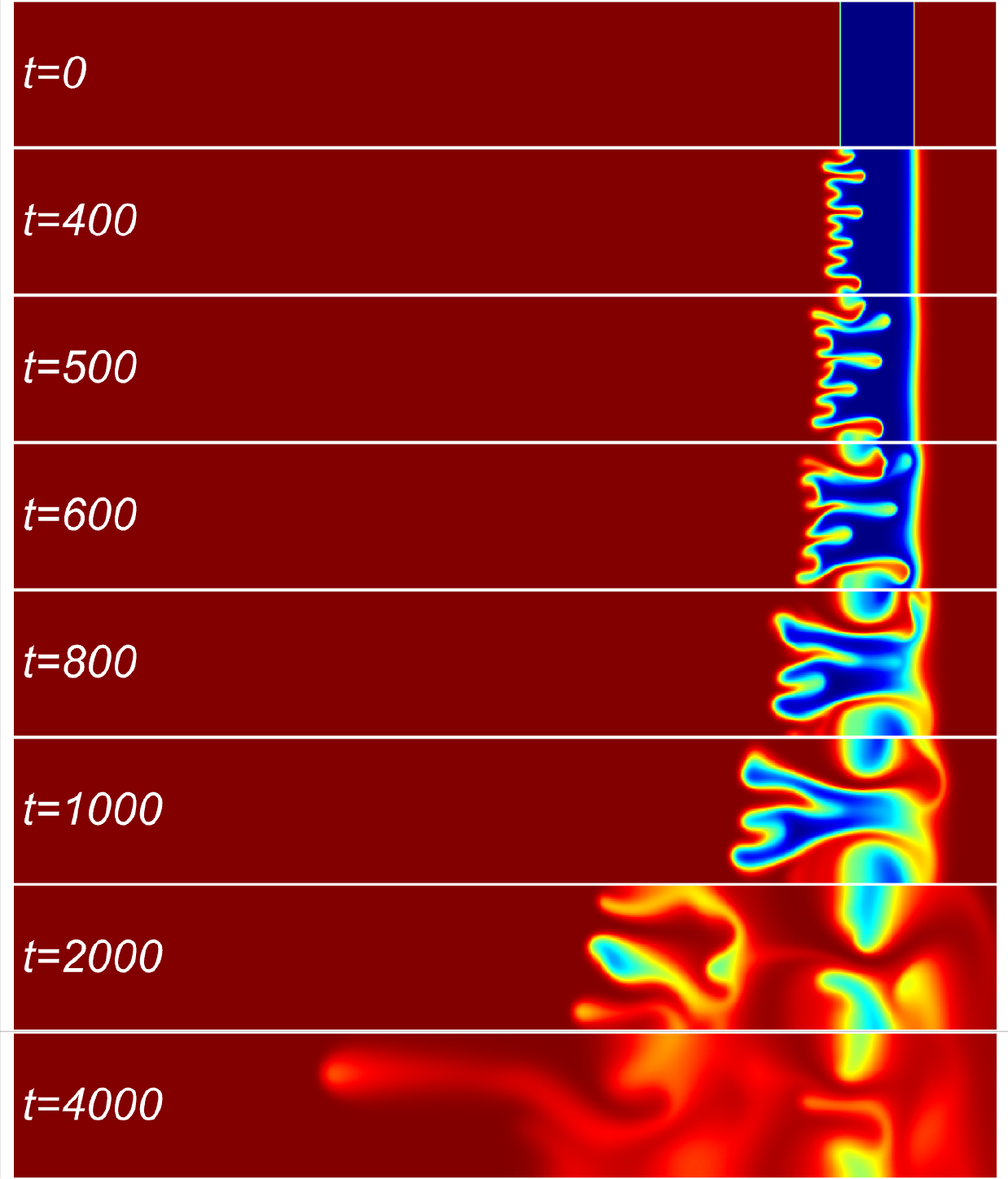}
    \\[0.2cm]
    \includegraphics[width=0.5\textwidth]{figures/colorbar.pdf}
    \caption{Spatio-temporal distribution of a more viscous slice of width $l = 256$ for $R = 3$ corresponding to (a) Type-I, (b) Type-II, and (c) Type-III boundary conditions. Other parameters are the same as in the figure \ref{Fig:Type-I}.} 
    \label{Fig:R3}
\end{figure}

In this section, we investigate the onset of viscous fingering, pattern formation, and the long-time behavior affected by different boundary conditions. Figure \ref{Fig:R3} depicts the evolution of a more viscous finite sample of width $l = 256$, for $R = 3$ and under different boundary conditions. It is noted that the number of fingers is not influenced by the boundary conditions. Although the fingering dynamics at the initial stages are discernibly indistinguishable, as time progresses, the patterns are significantly different between the three different boundary conditions. In all the cases, at $t \approx 600$ fingers reach the stable frontal interface, which acts as a barrier to the forward propagating fingers. The first instant of interaction of the fingers originated from the unstable interface with the stable interface is called the breakthrough time and is denoted as $t_{bk}$. Forward fingers carry the less viscous fluid in the flow direction, whereas the backward fingers are those that move in the upstream direction, carrying the high viscous fluid \cite{mishra2008differences}. Figures \ref{Fig:R3}(a) and \ref{Fig:R3}(b) depict that for the Type-I and Type-II boundary conditions, the stable interface of the sample restricts the growth of the advancing fingers and reorients the direction of the fingers. Contrary to these boundary conditions, Type-III boundary conditions exhibit different dynamics post-breakthrough time $t_{bk}$. In this case, the forward-moving fingers are able to penetrate the stable frontal interface. Although these fingers distort the stable interface, their reduced strength -- attributed to the dilution of the concentration over time prevents them from further fingering or rapid spreading. Nevertheless, the sample still moves slightly further compared to the other two cases governed by diffusive spreading. Though the reoriented fingers move steadily in the upstream direction, quite a different scenario is observed (see figure \ref{Fig:R3}(c)). 
 
\begin{figure}
	\centering
	(a) \hspace{1.9 in} (b) \hspace{1.9 in} (c) \\ 
	\includegraphics[width=0.32\textwidth]{figures/fig4b.pdf}
	\includegraphics[width=0.32\textwidth]{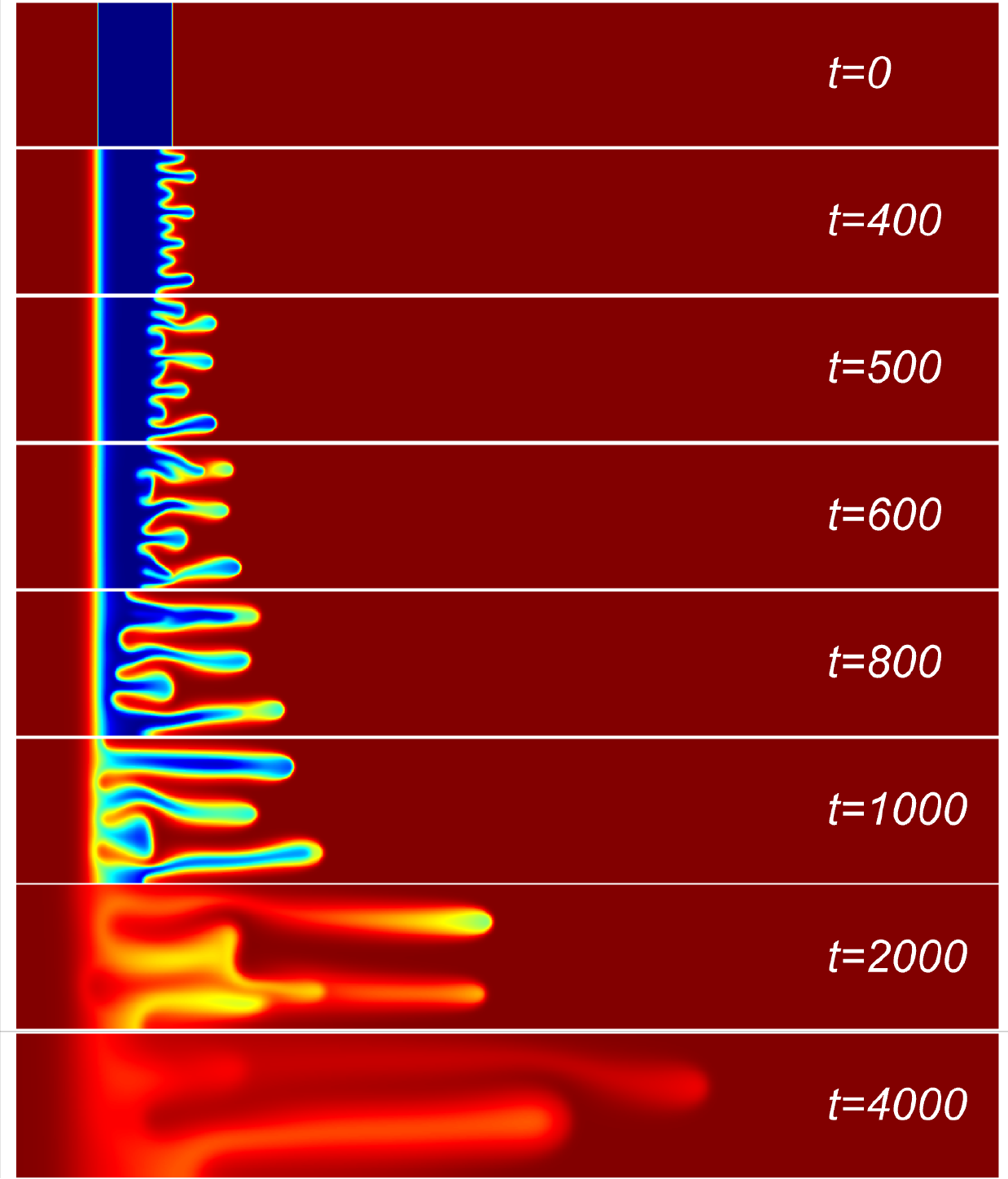}
	\includegraphics[width=0.32\textwidth]{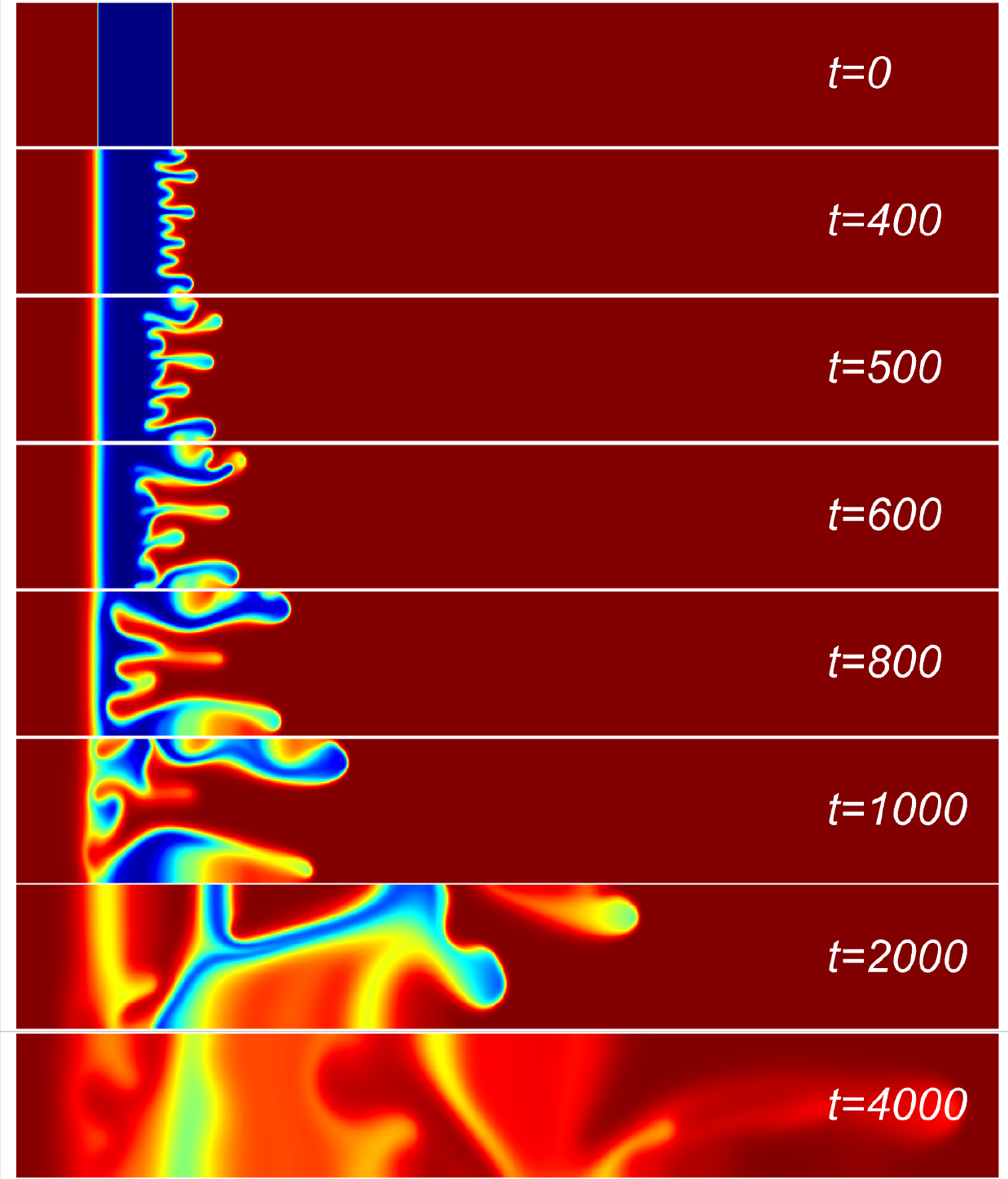}
	\\[0.2cm]
	\includegraphics[width=0.5\textwidth]{figures/colorbar.pdf}
	\caption{Spatio-temporal distribution of a less viscous slice of width $l = 256$ for $R = -3$ corresponding to (a) Type-I, (b) Type-II, and (c) Type-III boundary conditions. Other parameters are the same as in the figure \ref{Fig:Type-I}.} 
	\label{Fig:R_3}
\end{figure}

Figure \ref{Fig:R_3} depicts the deformation dynamics of a less viscous finite sample for the three different types of boundary conditions. All the parameters are the same as in figure \ref{Fig:R3} except $R = -3$. The values of the log-mobility ratio $R$ are so chosen that the viscosity ratio of the unstable interface remains the same in the less and more viscous samples. For a less viscous sample, the forward fingers propagate in the downstream direction without facing any barrier. Instead, the backward propagating fingers, originating from the frontal interface, interact with the stable rear interface, leading to the breakthrough. As expected, until the breakthrough, fingering dynamics at the unstable interfaces are identical for $R = -3$ (frontal interface) and $R = 3$ (rear interface). After the breakthrough time, the reoriented fingers move in the downstream direction. Unlike for a more viscous sample, a stable interface does not deform for a less viscous sample, corresponding to the case of the Type-III boundary conditions. In summary, the onset of fingering and early-time dynamics does not depend strongly on the choice of the boundary conditions. 
% \textcolor{red}{Furthermore, it is observed that an early breakthrough is encountered due to viscous fingering in a less viscous sample as compared to a more viscous sample, irrespective of the choice of the boundary conditions.} 
Post-breakthrough dynamics vary with the choice of the boundary conditions. 

To better understand the spreading and mixing of the sample, we resort to the transversely averaged concentration profiles and their quantitative properties discussed in \S \ref{subsec:avg_conc} - \S \ref{subsec:moments}. 

%%%%%%%%%%%%%%%%%%%%%%%%%%%%%%%%%%%%%%%%%%%%%%%%%%%%%%%%%%%%%%%%%%%%%%%%
%  R=-3
%%%%%%%%%%%%%%%%%%%%%%%%%%%%%%%%%%%%%%%%%%%%%%%%%%%%%%%%%%%%%%%%%%%%%%%%

\subsubsection{Mass balance}\label{subsec:mass-balance}

We denote the total mass of the solute and its normalized mass at time $t$ by $M(t)$, and $m(t)$, respectively, and they are defined as 
\begin{eqnarray}
\label{eq:Mass_area_integration}
M(t) = \int\limits_{0}^{L_y} \int\limits_{0}^{L_x}c(x,y,t){\rm d}x {\rm d}y, \quad \mbox{and} \quad m(t) = \frac{M(t)}{M(0)}.
\end{eqnarray}
Spatial average of the transport equation \eqref{eq:concentration} over the domain $\Omega$ and subsequent use of the Green's theorem yields
\begin{eqnarray}
    \frac{d}{dt}M(t) = \int_{\partial\Omega}\boldsymbol{J}\cdot\boldsymbol{n}{\rm d}s, 
    \label{eq:mass_balance}
\end{eqnarray}
where $\boldsymbol{J} = (D \boldsymbol{\nabla} c - \boldsymbol{u} c)$ is the total flux, and $\partial \Omega = \Gamma_1 \cup \Gamma_2 \cup \Gamma_3 \cup \Gamma_4 = \{x=0\}\cup\{y=L_y\}\cup\{x=L_x\}\cup\{y=0\}$ is the boundary of the domain $\Omega$. 
Integrating equation \eqref{eq:mass_balance} we obtain
\begin{equation}
	\label{eq:M_tilde}
	M(t) = M(0) + \int_{0}^{t} \int_{\partial\Omega}\boldsymbol{J}\cdot\boldsymbol{n}{\rm d}s {\rm d}\tau,
\end{equation}
such that 
\begin{equation}
\label{eq:m_tilde}
    m(t) = 1 + \int_0^t \int_{\partial \Omega} \frac{\boldsymbol{J}\cdot\boldsymbol{n}}{M(0)} {\rm d}s {\rm d}\tau. 
\end{equation}
We discussed below the effects of the three boundary types on the mass balance. 

First, consider Type-I boundary conditions. In this case, using the prescribed boundary conditions on the left and right boundaries, $\Gamma_1$ and $\Gamma_3$, we obtain $\boldsymbol{J} \cdot \hat{n} = 0$. Periodic boundary conditions in the transverse direction lead to periodic $\boldsymbol{J} \cdot \hat{n}$ in the transverse direction, i.e., $\left. \boldsymbol{J}\right\rvert_{\Gamma_2} = \left. \boldsymbol{J}\right\rvert_{\Gamma_4}$. Consequently, the contributions to the influx and outflux from the top and bottom boundaries cancel each other out.

\begin{figure}
	\centering
	(a) \hspace{3.2 in} (b) \\ 
	\includegraphics[width=0.495\textwidth]{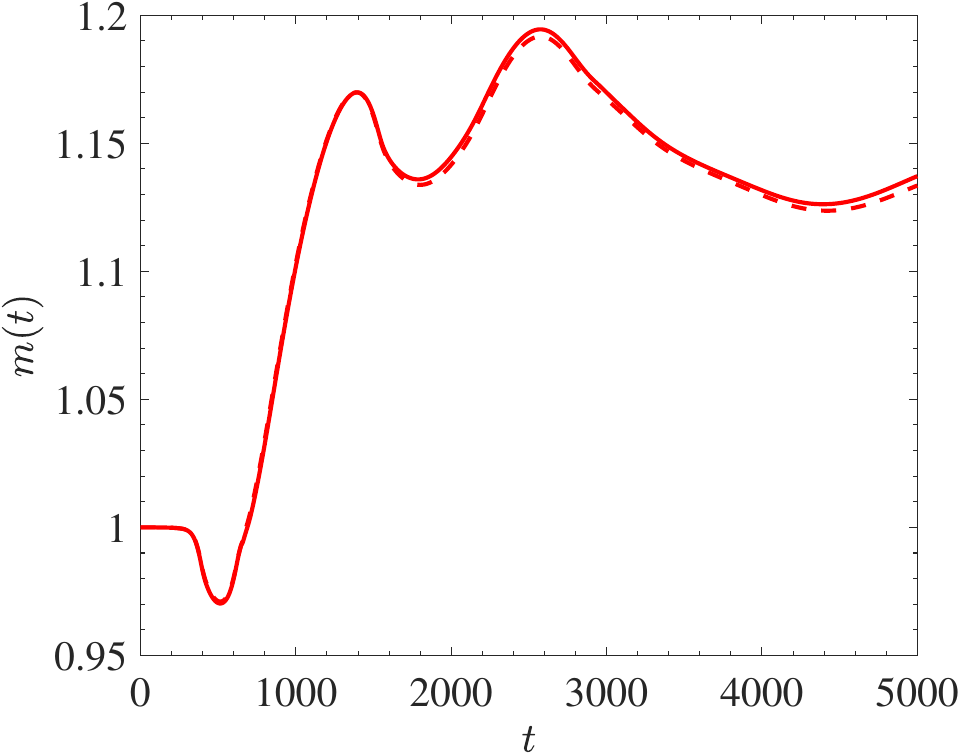} 
	\includegraphics[width=0.495\textwidth]{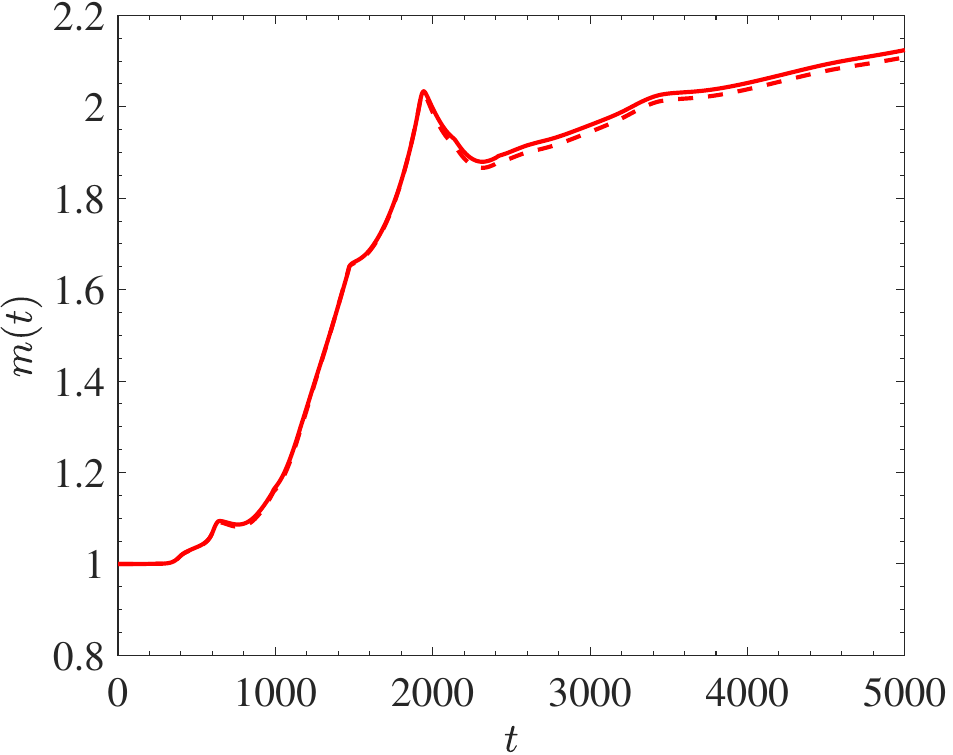}
	\caption{Temporal evolution of the normalized mass $m(t)$ for (a) $R = 3,$ (b) $R = -3$ corresponding to Type-III boundary conditions. The solid line represents the mass accumulating through area integration, whereas the dashed line represents the mass from flux calculation. Other parameters are the same as in the figure \ref{Fig:Type-I}.}
	\label{Fig:Mass_neumann}
\end{figure}

On the other hand, for the case of Type-II boundary conditions, the no flux condition of solute concentration on the boundaries, and $\psi=0$, in the transverse direction, causes the flux to vanish across the boundaries. Therefore, in both these cases, we have 
\begin{equation}
%	\label{eq:m_tilde}
\frac{d}{dt}M(t) = \int_{\partial\Omega}\boldsymbol{J}\cdot\hat{n}{\rm d}s = 0,
\end{equation}
implying that total mass is conserved. Next, we consider the Type-III boundary conditions for which we calculate the mass using equations \eqref{eq:Mass_area_integration} and \eqref{eq:M_tilde}. Numerical computations of the normalized solute mass using equations \eqref{eq:Mass_area_integration} and \eqref{eq:m_tilde} yield a significant deviation from the initial mass (see figure \ref{Fig:Mass_neumann}). While mass decreases initially for $R > 0$, it eventually increases as time progresses. On the other hand, for $R < 0$, although the evolution of the solute mass is non-monotonic, it never goes below the initial mass such that $m(t) \geq 1$, $\forall t \geq 0$. The change of mass in the system is attributed to the net flux across the domain boundaries. As demonstrated in figure \ref{Fig:Mass_neumann}, close agreements between the normalized mass computed using equations \eqref{eq:Mass_area_integration} and \eqref{eq:m_tilde} ensure the accuracy of the numerical methods, besides explaining the physics of mass balance. 

%%%%%%%%%%%%%%%%%%%%%%%%%%%%%%%%%%%%%%%%%%%%%%%%%%%%%%%%%%%%%%%%%%%%%%%%
% figure- [Total mass balance under Type-III condition (R = 3, -3)]
%%%%%%%%%%%%%%%%%%%%%%%%%%%%%%%%%%%%%%%%%%%%%%%%%%%%%%%%%%%%%%%%%%%%%%%%

In summary, from numerical simulation results, we found that for the Type-I and II boundary conditions, solute mass remains conserved. However, for the Type-III boundary conditions, the solute mass increases due to flux across the transverse boundaries. As in later times, as solute mass increases, the concentration gradient may become higher or persist over a longer time interval, leading to stronger instability. 

\subsubsection{Transverse-averaged concentration}\label{subsec:avg_conc}

\begin{figure}
		\centering
		(a) \hspace{3.2 in} (b) \\ 
		\includegraphics[width=0.495\textwidth]{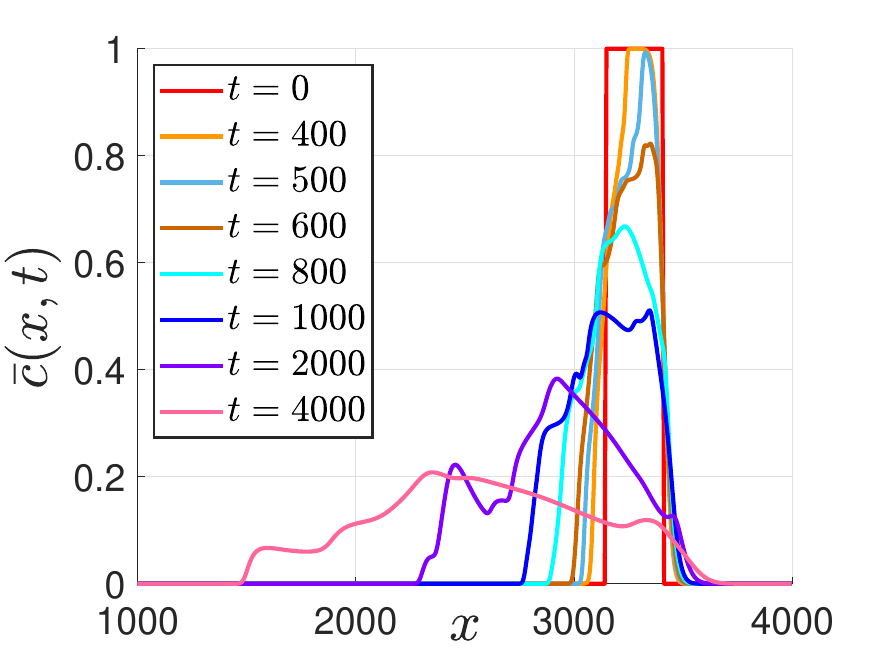} 
		\includegraphics[width=0.495\textwidth]{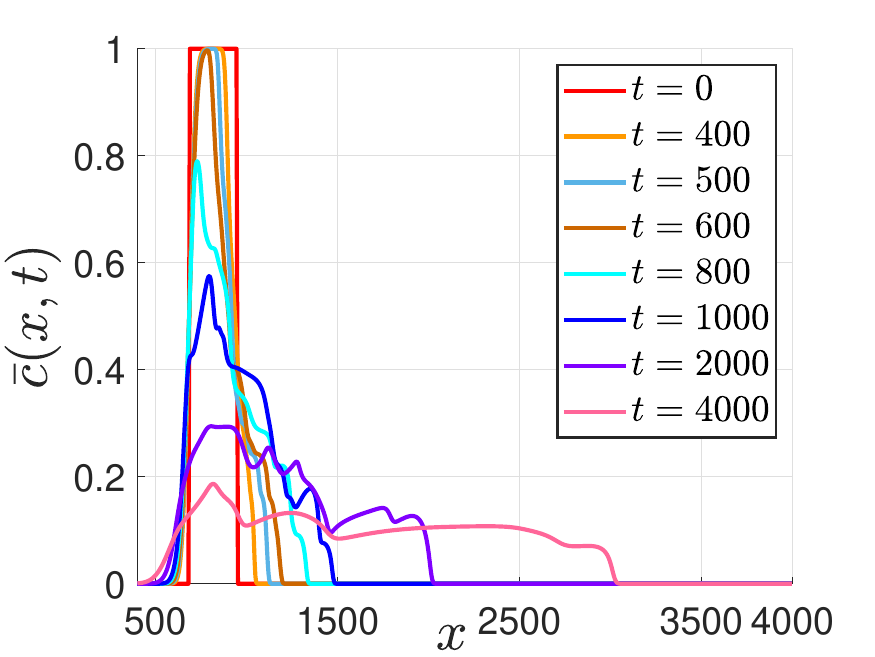}
		\caption{Transverse-averaged concentration profile for (a) $R = 3,$ and (b) $R = -3$ corresponding to Type-I boundary conditions. Other parameters are the same as in the figure \ref{Fig:Type-I}.}
		\label{Fig:Periodic_TAC}
	\end{figure}

\begin{figure}
		\centering
		(a) \hspace{3.2 in} (b) \\ 
		\includegraphics[width=0.495\textwidth]{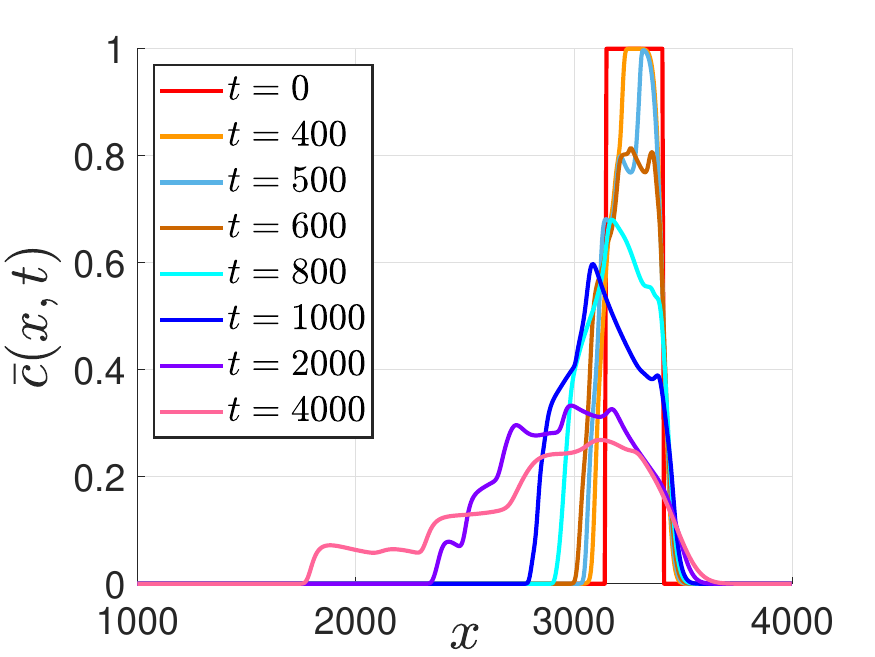} 
		\includegraphics[width=0.495\textwidth]{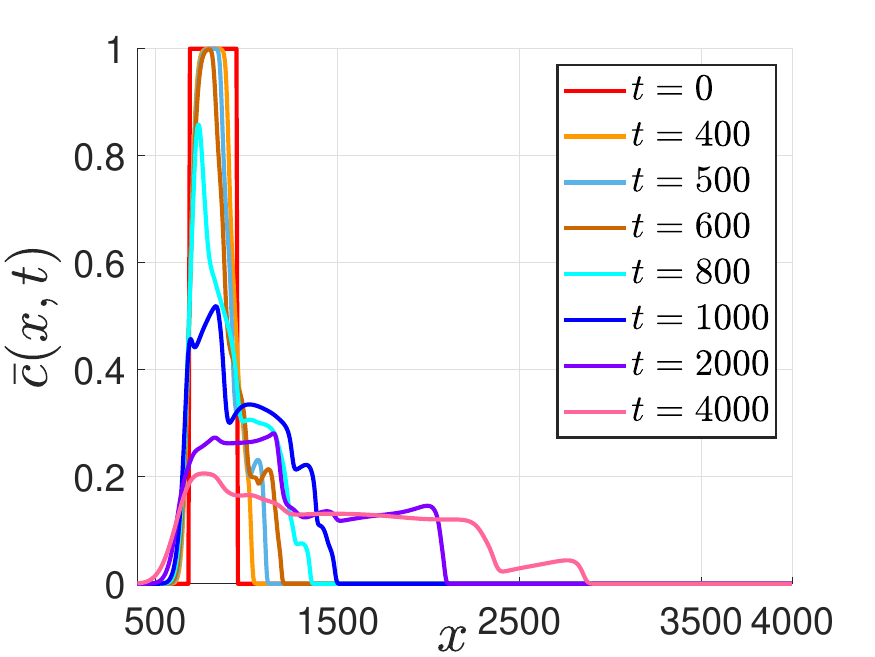}
		\caption{Transverse-averaged concentration profile for (a) $R = 3,$ and (b) $R = -3$ corresponding to Type-II boundary conditions. Other parameters are the same as in the figure \ref{Fig:Type-I}.}
		\label{Fig:Dirichlet_TAC}
	\end{figure}

\begin{figure}
	\centering
	(a) \hspace{3.2 in} (b) \\ 
	\includegraphics[width=0.495\textwidth]{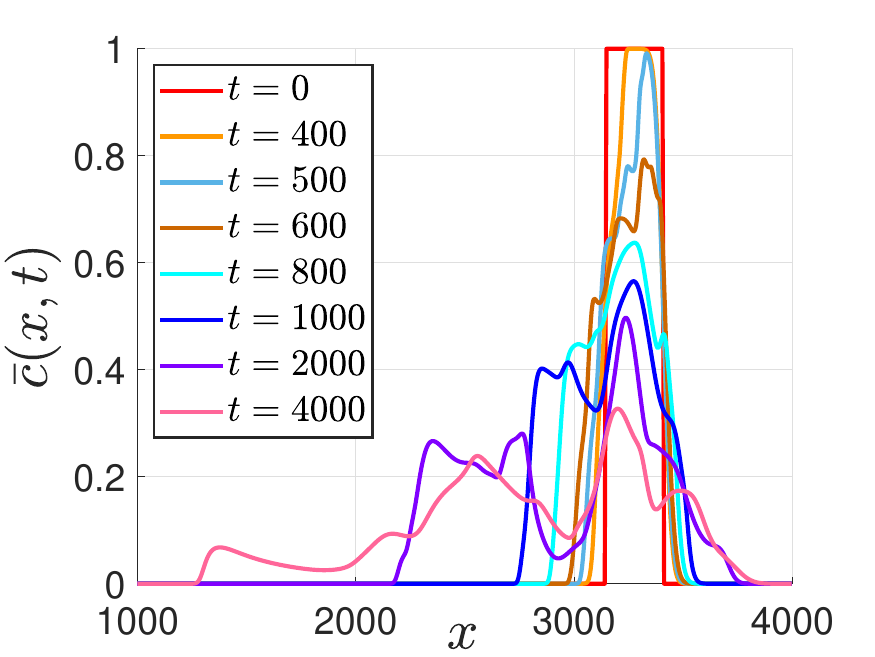} 
	\includegraphics[width=0.495\textwidth]{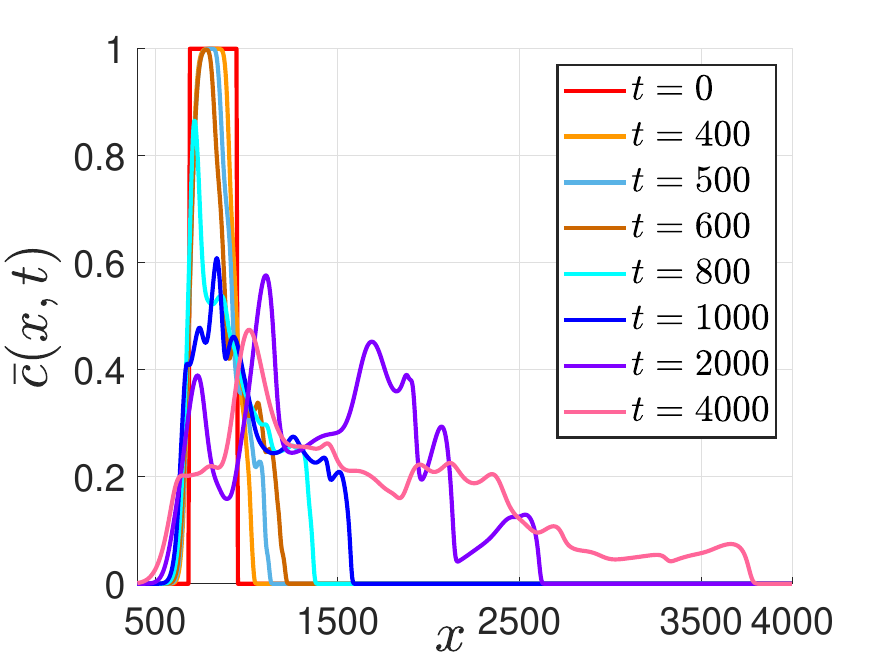}
	\caption{Transverse-averaged concentration profile for (a) $R = 3,$ and (b) $R = -3$ corresponding to Type-III boundary conditions. Other parameters are the same as in the figure \ref{Fig:Type-I}.}
	\label{Fig:Neumann_TAC}	
\end{figure}

	%%%%%%%%%%%%%%%%%%%%%%%%%%%%%%%%%%%%%%%%%%%%%%%%%%%%%%%%%%%%%%%%%%%%%%%%
The transverse-averaged concentration profile defined in equation \eqref{eq:trans_avg} provides the information of the concentration profile along the domain. Figures \ref{Fig:Periodic_TAC}, \ref{Fig:Dirichlet_TAC}, and \ref{Fig:Neumann_TAC} depict $\bar{c}(x,t)$ corresponding to the Type-I, II, and III boundary conditions, respectively. As discussed in \S \ref{sec:slice_viscosity_matched}, for $ R = 0 $, transverse-averaged concentration profiles exhibit a unimodal distribution, a feature of stable displacement of a finite sample. Therefore, a departure from a unimodal to a multi-modal distribution corresponds to fingering instability. 

Before the breakthrough time $t_{bk}$, i.e., the time when the advancing finger front interacts with the diffusive front, the frontal (rear) interface remains stable for $R > 0$ ($R < 0$), which can be described using the error function -- a feature of diffusive spreading. However, for $t > t_{bk}$, the fate of the stable interface depends on the choice of the boundary conditions. For Type-I and II boundary conditions, the stable interface acts as a barrier to the finger propagation and penetration. As a result, the stable interface maintains its error function profile, as depicted in figure \ref{Fig:Periodic_TAC} and \ref{Fig:Dirichlet_TAC}. Contrary to these, for Type-III boundary conditions, the dynamics of $\bar{c}(x, t)$ differ significantly post the breakthrough time. In particular, for $R = 3$, after the breakthrough time, fingers can push further downstream, causing the initially stable interface to lose its diffusive profile (see figure \ref{Fig:Neumann_TAC}(a)). As discussed in \S \ref{subsec:mass-balance}, mass flux through the transverse boundaries enhances the solute mass that continuously feeds to the concentration gradient in the case of Type-III boundary conditions. This leads to an increase in the solute concentration gradient. Thus, the viscosity contrast between the displacing and defending fluids is higher in this case compared to the other two cases, leading to a stronger fingering instability. Although mass enhancement for $R = -3$ is larger than that of $R = 3$, as the fingers moving towards the stable interface are against the background flow for $R = -3$, the upstream fingers fail to break through the stable interface in the latter, unlike the former (see figure \ref{Fig:Neumann_TAC}(a-b)). Nevertheless, at the stable interface, the sample spreads more in the case of Type-III conditions compared to the other two cases of boundary conditions, which follow a diffusive trend. This enhanced spreading can be measured quantitatively using the mixing length discussed in the next section. 

	%%%%%%%%%%%%%%%%%%%%%%%%%%%%%%%%%%%%%%%%%%%%%%%%%%%%%%%%%%%%%%%%%%%%%%%%
	%%%%%%%%%%%%%%%%%%%%%%%%%%%%%%%%%%%%%%%%%%%%%%%%%%%%%%%%%%%%%%%%%%%%%%%%
	% figure- [Transverse-averaged profile under Type-II condition (R = 3, -3)]
	%%%%%%%%%%%%%%%%%%%%%%%%%%%%%%%%%%%%%%%%%%%%%%%%%%%%%%%%%%%%%%%%%%%%%%%%	
%%%%%%%%%%%%%%%%%%%%%%%%%%%%%%%%%%%%%%%%%%%%%%%%%%%%%%%%%%%%%%%%%%%%%%%%
% figure- [Transverse-averaged profile under Type-III condition (R = 3, -3)]
%%%%%%%%%%%%%%%%%%%%%%%%%%%%%%%%%%%%%%%%%%%%%%%%%%%%%%%%%%%%%%%%%%%%%%%%

\subsubsection{Mixing length} \label{sec:mixing_length}

For the case of a finite-width slice, as the two interfaces give rise to forward or backward fingers depending upon the sign of $R$, we define two mixing lengths. The forward mixing length is defined as the distance between the initial position of the frontal interface of the slice to the point in the downstream direction up to $\bar{c}(x,t) = 0.01$, and is denoted by $L_d^+$. Similarly, the backward mixing length is defined as the distance between the initial position of the rear interface of the slice to the point in the upstream direction where $\bar{c}(x,t) = 0.01$ and is denoted by $L_d^-$. Mixing length for stable interface grows $\propto \sqrt t$. 

%%%%%%%%%%%%%%%%%%%%%%%%%%%%%%%%%%%%%%%%%%%%%%%%%%%%%%%%%%%%%%%%%%%%%%%%
% figure- [Mixing length comparison (R = 3)]
%%%%%%%%%%%%%%%%%%%%%%%%%%%%%%%%%%%%%%%%%%%%%%%%%%%%%%%%%%%%%%%%%%%%%%%%
	\begin{figure}
		\centering
		(a) \hspace{3.2 in} (b) \\ 
		\includegraphics[width=0.495\textwidth]{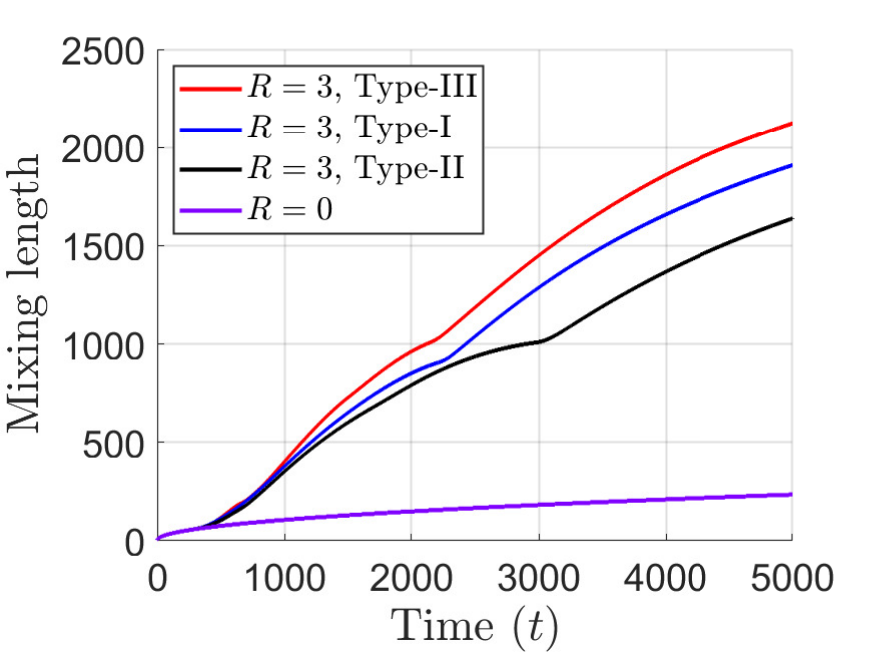} 
		\includegraphics[width=0.495\textwidth]{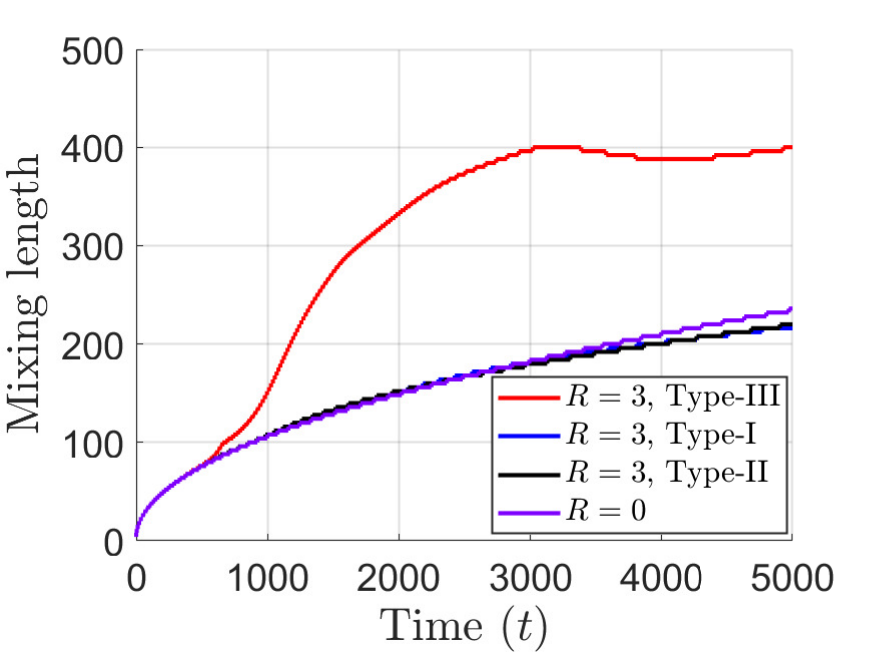}
		\caption{Comparison of mixing lengths (a) rear unstable $(R > 0)$ interface (backward mixing length, $L_d^-$) and (b) frontal stable interface (forward mixing length, $L_d^+$) for $R = 3$ corresponding to different types of boundary conditions. The stable diffusive mixing length is represented by $R = 0$. It clearly depicts the domination of mixing length for Type-III boundary conditions.}
		\label{Fig:R3_mixing}
	\end{figure} 
    
%%%%%%%%%%%%%%%%%%%%%%%%%%%%%%%%%%%%%%%%%%%%%%%%%%%%%%%%%%%%%%%%%%%%%%%%
% figure- [Mixing length comparison (R = -3)]
%%%%%%%%%%%%%%%%%%%%%%%%%%%%%%%%%%%%%%%%%%%%%%%%%%%%%%%%%%%%%%%%%%%%%%%%

\begin{figure}
	\centering
	(a) \hspace{3.2 in} (b) \\ 
	\includegraphics[width=0.495\textwidth]{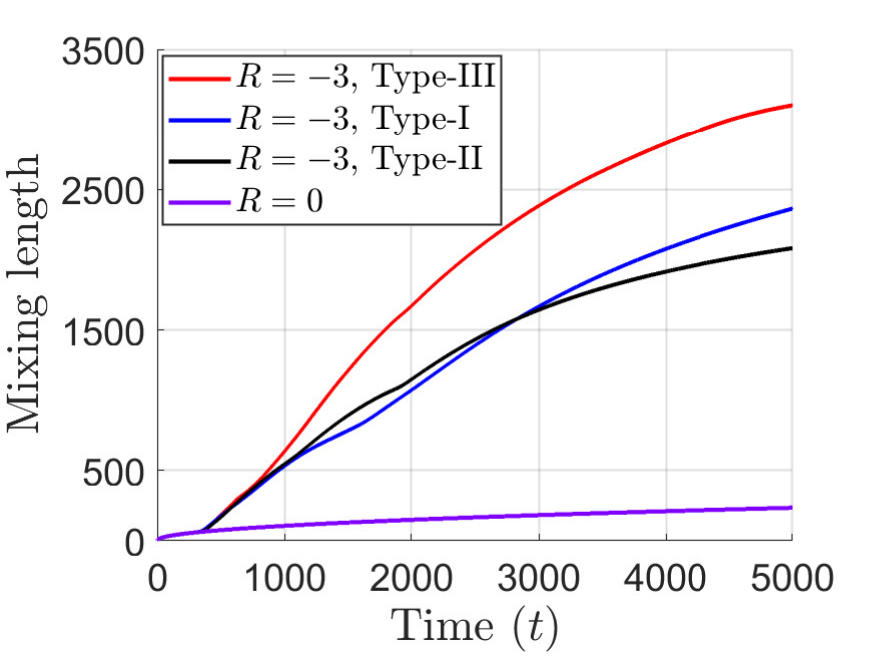} 
	\includegraphics[width=0.495\textwidth]{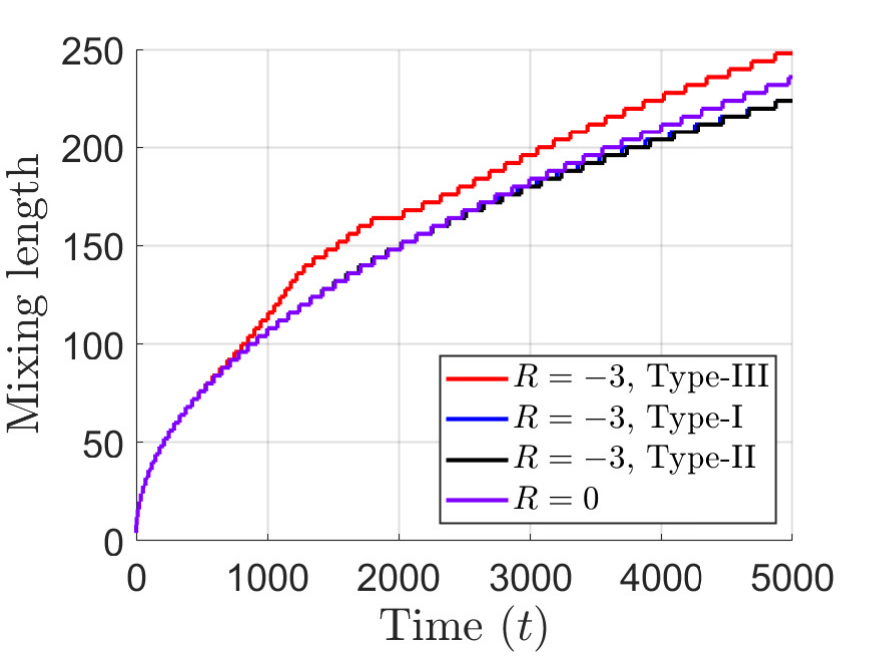}
	\caption{Comparison of mixing lengths (a) frontal unstable $(R < 0)$ interface (forward mixing length, $L_d^+$) and (b) rear stable interface (backward mixing length, $L_d^-$) for $R = -3$ corresponding to the different types of boundary conditions. $R = 0$ corresponds to the diffusive mixing length.}
	\label{Fig:R_3_mixing}
\end{figure} 

Figure \ref{Fig:R3_mixing}(a) illustrates the evolution of $L_d^-$ for $R = 3$ with different boundary conditions. It is observed that initially $L_d^-$ follows a diffusive evolution ($\propto \sqrt{t}$). Subsequently, they depart from $\sqrt{t}$ behaviour and grow rapidly, characteristics of viscous fingering instability. The instant when $L_d^{-}$ transitions to rapid growth from $\sqrt{t}$ behaviour, is characterized as the onset of viscous fingering and is denoted by $t_{on}$. 

Figures \ref{Fig:R3_mixing}(a) and \ref{Fig:R_3_mixing}(a) depict that, for the Type-III boundary conditions, mixing is more than in the remaining cases. As mentioned earlier, the Type-III boundary condition allows mass enhancement in the system, enabling the maintenance of a higher concentration gradient, which results in higher viscosity contrasts. Hence, it promotes stronger VF instability, leading to an enhanced mixing compared to the other two cases. Note that beyond $t_{on}$, mixing length increases with time before it experiences an instantaneous decrease followed by a subsequent increase in time. This temporal decrease is caused by a dilution of the advancing finger, and the temporal increase is again due to a fingering growth of the adjacent finger. As multiple fingers develop and compete with each other, one or more advancing fingers fade away, and the adjacent finger(s) dominate(s) in the mixing process. 

%%%%%%%%%%%%%%%%%%%%%%%%%%%%%%%%%%%%%%%%%%%%%%%%%%%%%%%%%%%%%%%%%%%%%%%%
% figure- [Mixing length comparison at untsbale interface]
%%%%%%%%%%%%%%%%%%%%%%%%%%%%%%%%%%%%%%%%%%%%%%%%%%%%%%%%%%%%%%%%%%%%%%%%
\begin{figure}
	\centering
	(a) \hspace{3.1 in} (b) \\
	\includegraphics[width=0.495\textwidth]{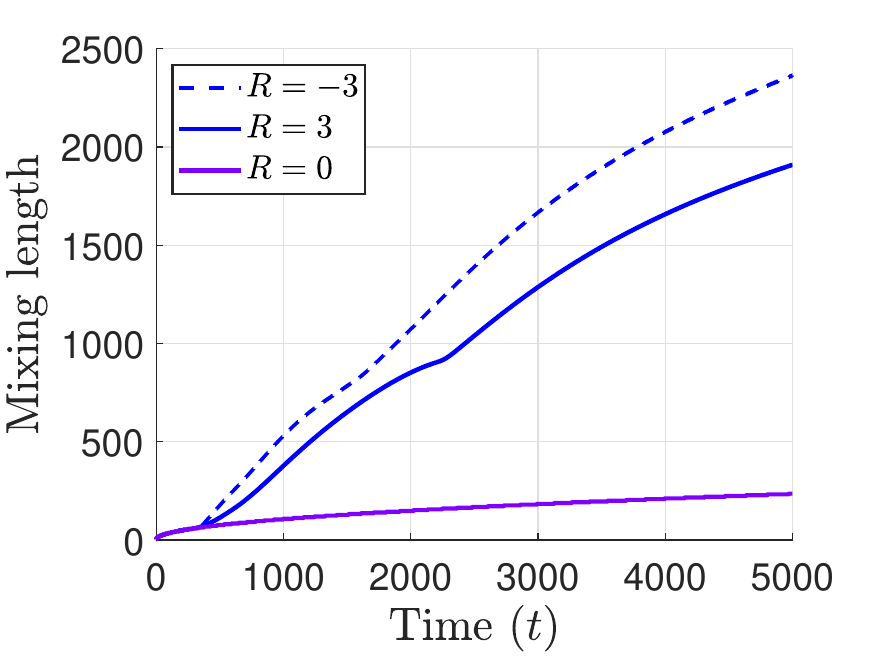}
	\includegraphics[width=0.495\textwidth]{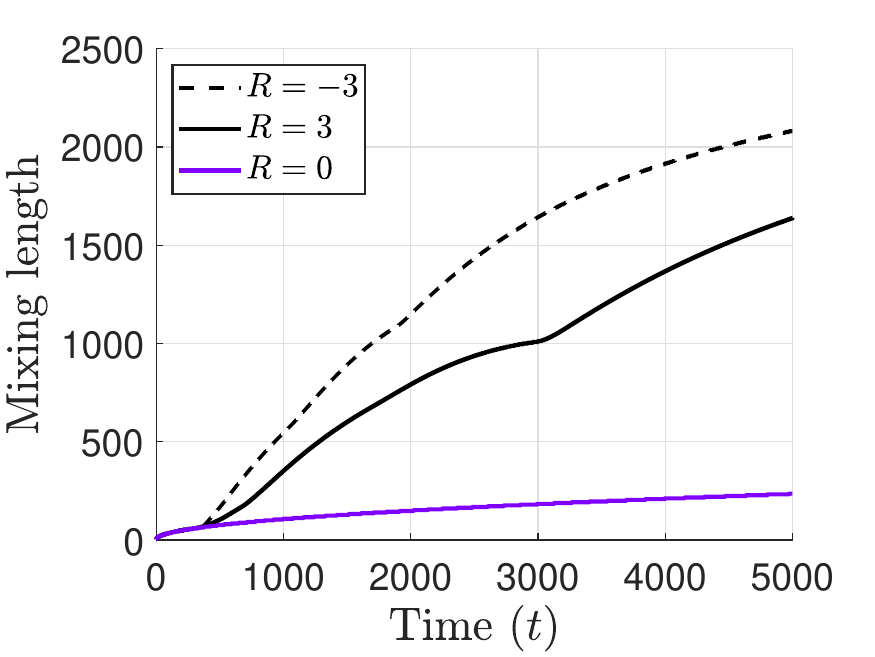} \\
	(c) \\
	\includegraphics[width=0.5\textwidth]{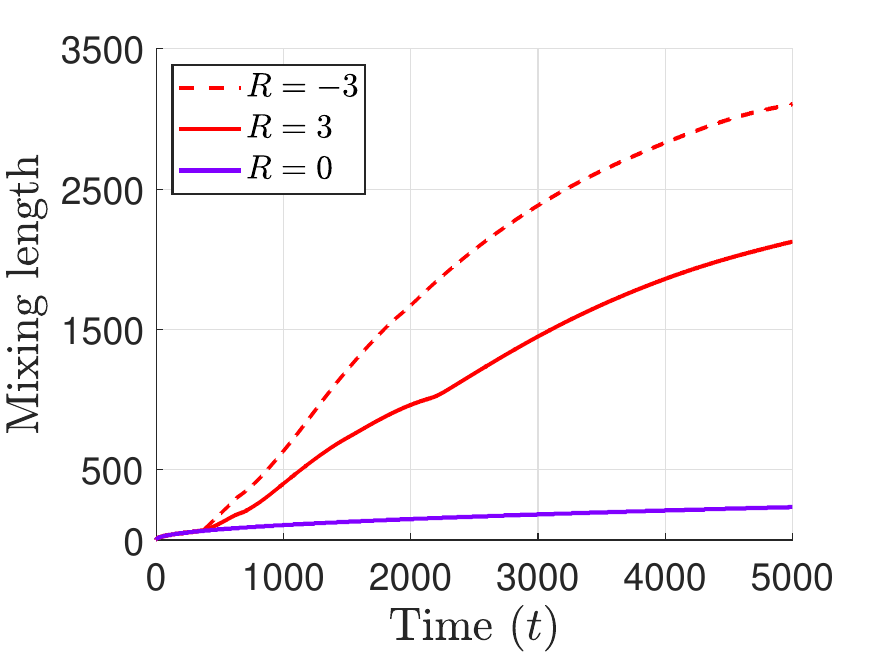} 
	\caption{Comparison of mixing lengths at the rear unstable interface for $R = 3$ (backward mixing length, $L_d^-$) and at the frontal unstable interface for $R = -3$ (forward mixing length, $L_d^+$) for different boundary conditions: (a) Type-I, (b) Type-II, and (c) Type-III. Diffusive mixing is represented by $R = 0$.}
	\label{Fig:Mixing_all}
\end{figure} 

On the other hand, at the stable interface, mixing length evolves as $\propto \sqrt{t}$ and remains consistent for all three cases until the breakthrough time $t_{bk}$. Subsequently, distinct behavior emerges depending on boundary conditions. Type-III boundary conditions allow mass enhancement in the system, leading to stronger VF instability. This enhancement breaks the stable barrier and results in larger mixing than the pure diffusion case ($R = 0$). Specifically, for $R = 3$, the advancing finger propagates farther through the stable interface, causing broader spreading than for $R = -3$ (compare between figure \ref{Fig:R3_mixing}(b) and figure \ref{Fig:R_3_mixing}(b)). 
%Figures \ref{Fig:R3_mixing}(b) and \ref{Fig:R_3_mixing}(b) confirm the broader spreading for $R=3$ than $R=-3$ in the case of Type-III condition, followed by $R=0$.	

In contrast, for the Type-I and II boundary conditions, for $t > t_{bk}$, fingers originating from the unstable interface are unable to penetrate the stable barrier and reorient the finger direction. For $R > 0$, the higher concentration near the stable interface moves in the upstream; while for $R < 0$, it moves downstream, bringing about a reduced concentration gradient near the stable interface, compared to the pure diffusion case. As a result, in the long-time behaviour, mixing length at the stable interface is slightly larger for $R = 0$ than for $R\neq 0$. 

Finally, we compare the mixing of the unstable interface of a more viscous sample ($R = 3$) with that of a less viscous sample ($R = -3$) for a specific choice of the boundary conditions. As forward fingers are moving in the downstream direction, and backward fingers are in the upstream direction, mixing length is larger for $R=-3$ than that for $ R = 3 $, in agreement with results shown in the figures \ref{Fig:Periodic_TAC}--\ref{Fig:Neumann_TAC} and \ref{Fig:Mixing_all}. 

\subsubsection{Interfacial length} \label{subsec:Interfacial_length}

Following earlier works \citep{mishra2008differences}, we define interfacial length as 
	\begin{equation}
		I(t) = \int\limits_{0}^{L_x}{\displaystyle\int\limits_{0}^{L_y}} |\nabla c|{\rm d}x{\rm d}y.
	\end{equation}
It measures the temporal variation of the concentration gradient across the domain, and captures the onset of the viscous fingering as well as the onset of the interaction between the front and rear interface of the slice. In the diffusive regime $(t<t_{on})$, interfacial length remains constant (equal to the twice the width of the domain, i.e., $2 L_y$) and with non-zero $R$, for $t>t_{on}$, fingers develops at the unstable interface, leading $I(t)$ to deviate from the constant value. 

%%%%%%%%%%%%%%%%%%%%%%%%%%%%%%%%%%%%%%%%%%%%%%%%%%%%%%%%%%%%%%%%%%%%%%%%
% figure- [Interfacial length for fix (R = 3, -3) under different 
%boundary condition]
%%%%%%%%%%%%%%%%%%%%%%%%%%%%%%%%%%%%%%%%%%%%%%%%%%%%%%%%%%%%%%%%%%%%%%%%
\begin{figure}
	\centering
	(a) \hspace{3.2 in} (b) \\ 
	\includegraphics[width=0.495\textwidth]{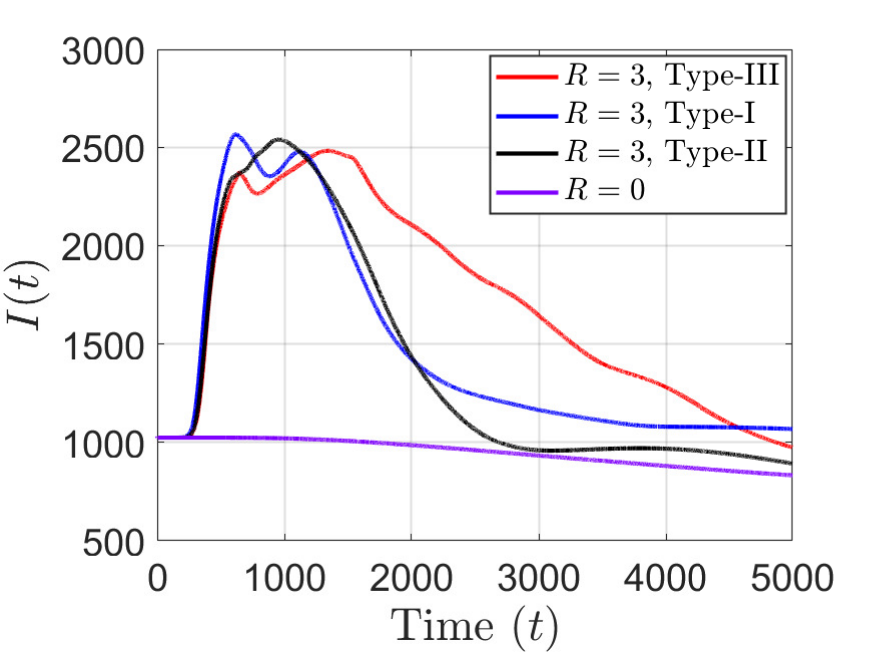} 
	\includegraphics[width=0.495\textwidth]{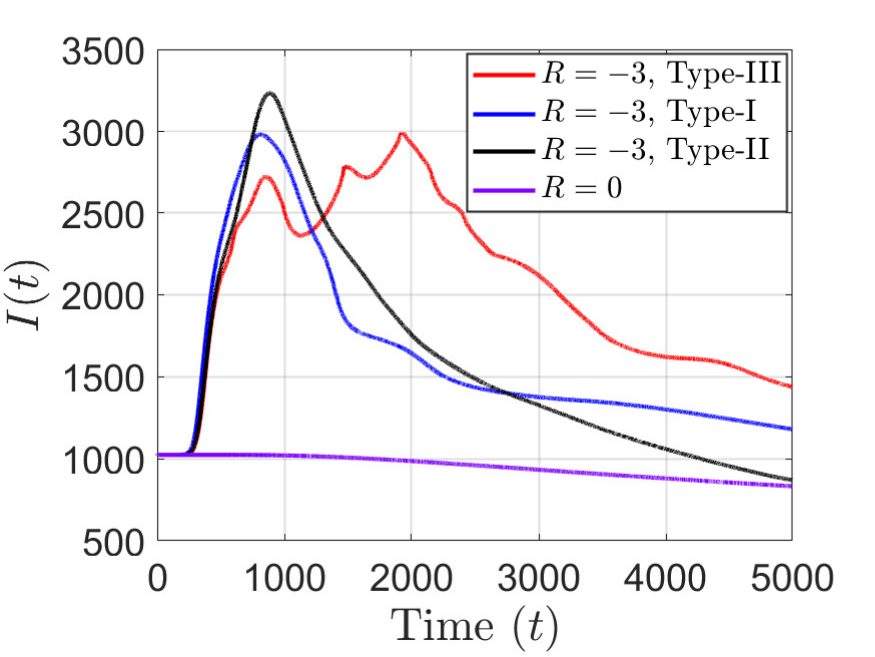}
	\caption{Temporal evolutions of the interfacial lengths for (a) $R = 3$ and (b) $R = -3$ corresponding to the different type of boundary conditions. Other parameters are the same as in the figure \ref{Fig:Type-I}.}
	\label{Fig:Interfacial}
\end{figure}

%%%%%%%%%%%%%%%%%%%%%%%%%%%%%%%%%%%%%%%%%%%%%%%%%%%%%%%%%%%%%%%%%%%%%%%%
% figure- [Interfacial length for fixed boundary condition with  
% different (R = 3, -3) ]
%%%%%%%%%%%%%%%%%%%%%%%%%%%%%%%%%%%%%%%%%%%%%%%%%%%%%%%%%%%%%%%%%%%%%%%%
\begin{figure}
	\centering
	(a) \hspace{3.1 in} (b) \\
	\includegraphics[width=0.495\textwidth]{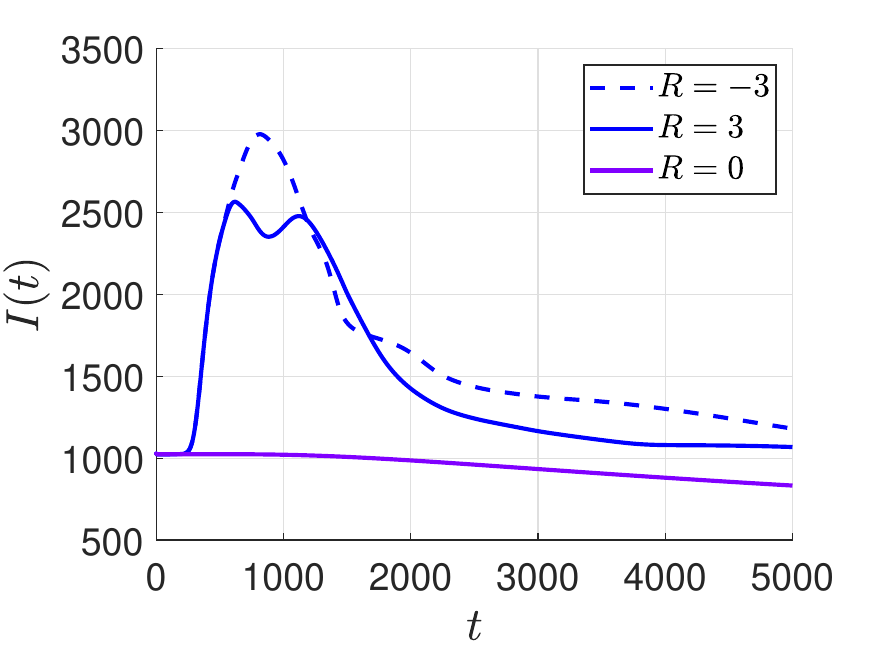}
	\includegraphics[width=0.495\textwidth]{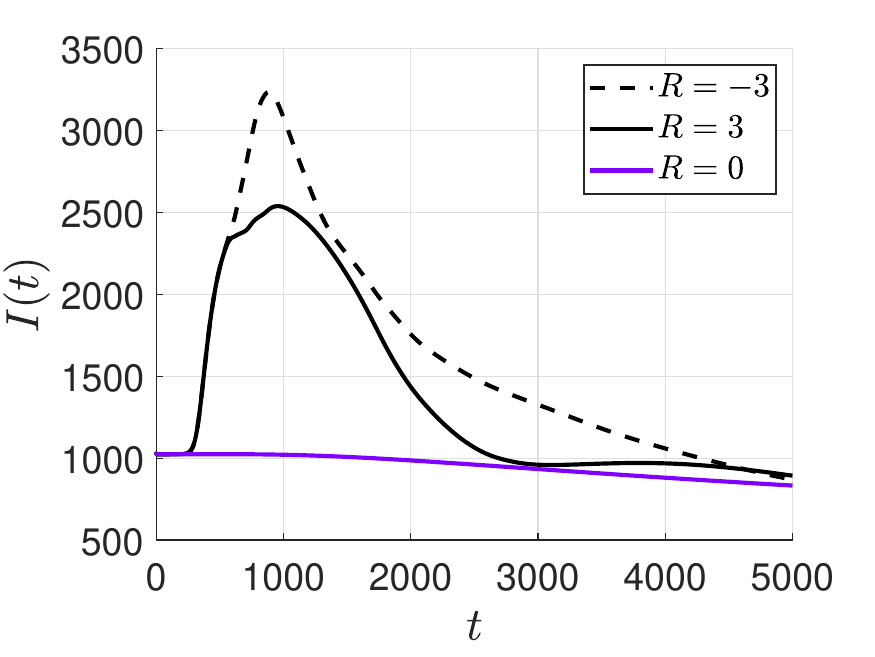} \\
	(c) \\
	\includegraphics[width=0.5\textwidth]{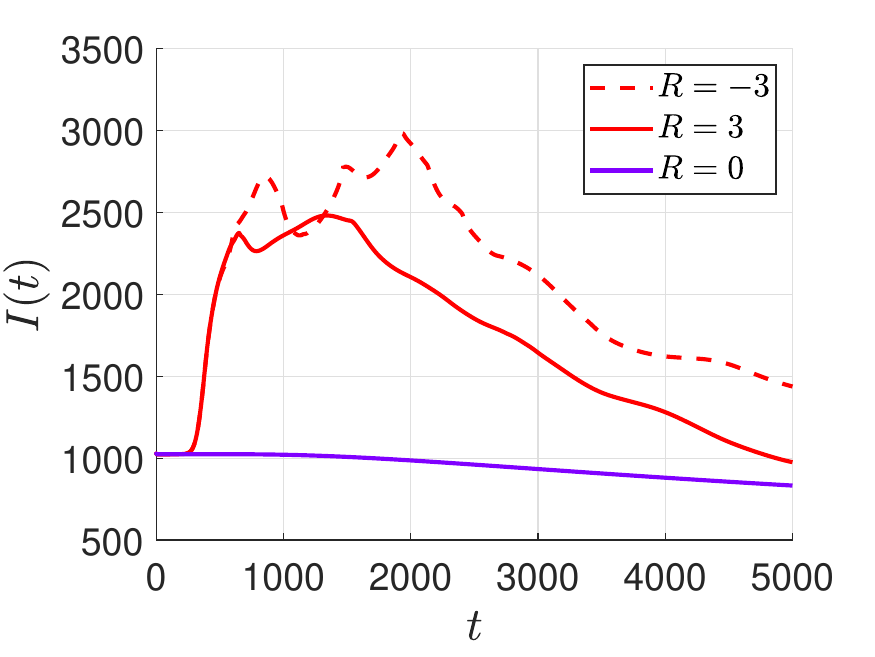} 
	\caption{Temporal evolution of interfacial lengths for $R = 3$ and $R = -3$ corresponding to (a) Type-I, (b) Type-II, and (c) Type-III boundary conditions. Diffusive interfacial length is represented by $R=0$. Other parameters are the same as in the figure \ref{Fig:Type-I}.}
	\label{Fig:Interfacial_all}
\end{figure} 

After the onset of VF $(t>t_{on})$, advancing fingers enhances the fluid-fluid interface, causing $I(t)$ to increase, as depicted in figure \ref{Fig:Interfacial}. However, the interaction of the fingers with the stable barrier leads to a reduction in the interfacial length. As discussed earlier, the spatio-temporal dynamics of the solute concentration until the breakthrough time $t_{bk}$ are indistinguishable for both $R=3$ and $R=-3$. Consequently, the evolution of the interfacial length $I(t)$ are the same for both more or less viscous samples until the breakthrough time of $R = 3$, (see in figure \ref{Fig:Interfacial_all}). This observation also explains why $t_{bk}$ is comparatively larger for $R=-3$ than for $R=3$, as the fingers have to move against the flow for $R=-3$. 

In all cases discussed here, a decline in $I(t)$ after the breakthrough (i.e., for $t > t_{bk}$) reflects a reduction in the concentration gradient. This decay appears either from the interaction between rear and frontal interfaces of the sample or from the merging of adjacent fingers, both of which diminish the available interfacial area. Under Type-I and Type-II boundary conditions, solute dilution progressively weakens the concentration gradients, resulting in a monotonic decay of $I(t)$ as demonstrated in \ref{Fig:Interfacial_all}(a) and (b).

By contrast, Type-III boundary conditions produce a markedly different evolution. As shown in \ref{Fig:Interfacial_all}(c), $I(t)$ exhibits a sequence of peaks and valleys after the breakthrough, driven by localized mass accumulation that persists or amplifies concentration gradients. While the following valleys are related to finger coalescence and the deterioration of the fluid-fluid interface, peaks reflect transient interface growth. The interplay between the enhancing and merging processes results in the non-monotonic behavior characteristic of the Type-III scenario.

\subsubsection{Moments} \label{subsec:moments}

%%%%%%%%%%%%%%%%%%%%%%%%%%%%%%%%%%%%%%%%%%%%%%%%%%%%%%%%%%%%%%%%%%%%%%%%
% figure- [Statistical first and second moments for different boundary 
%conditions with different (R = 3, -3) ]
%%%%%%%%%%%%%%%%%%%%%%%%%%%%%%%%%%%%%%%%%%%%%%%%%%%%%%%%%%%%%%%%%%%%%%%%
\begin{figure}
	\centering
	(a) \hspace{3.2 in} (b) \\ 
	\includegraphics[width=0.495\textwidth]{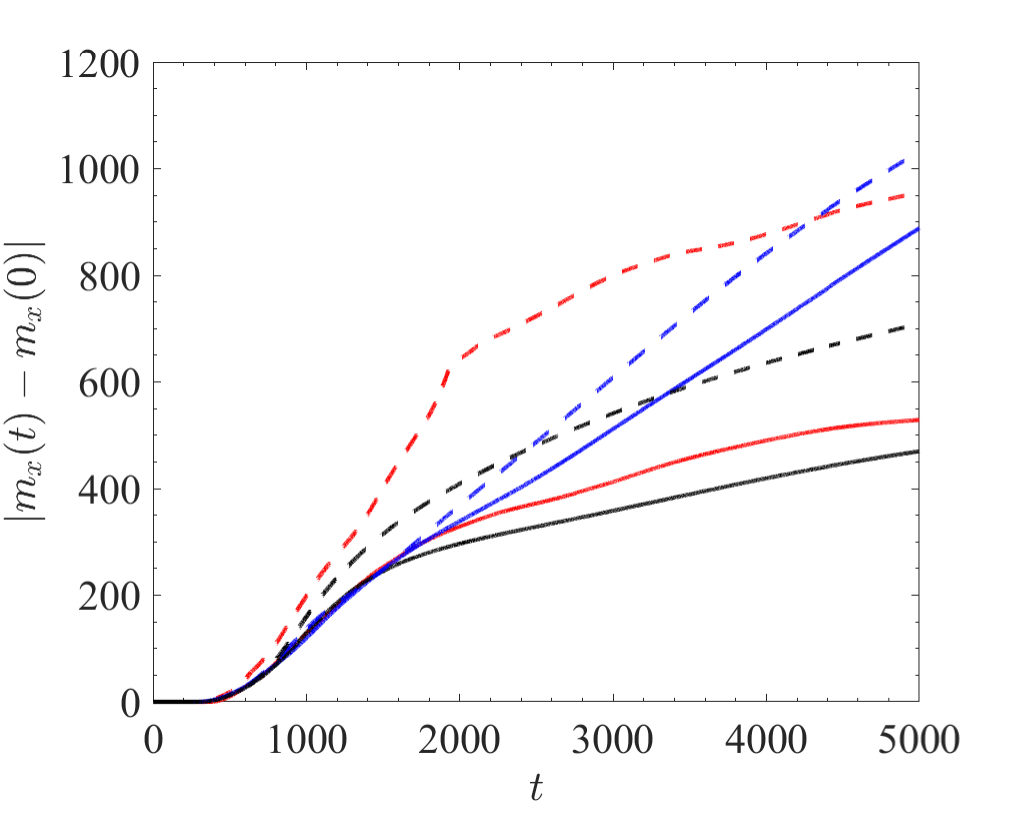} 
	\includegraphics[width=0.495\textwidth]{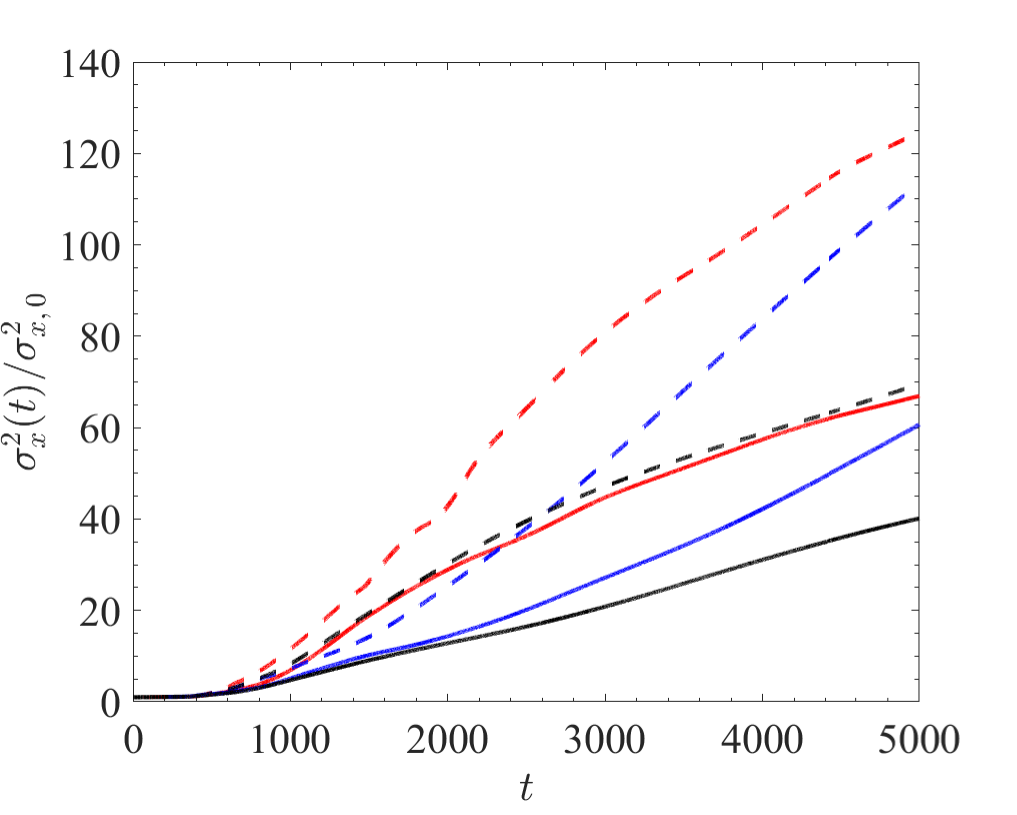}
	\caption{ Temporal evolution of (a) distance of the mean position from its initial value, (b) variance normalized by its initial value for different flow conditions: Type-I (blue), Type-II (black), and Type-III (red). It is observed that the mean position and the variance of a less viscous sample ($R = -3$, dashed lines) are higher than the corresponding more viscous sample ($R = 3$, solid lines).  Other parameters are the same as in the figure \ref{Fig:Type-I}.}
	\label{Fig:mean}
\end{figure} 
 \begin{figure}
	\centering
	\includegraphics[width=0.495\textwidth]{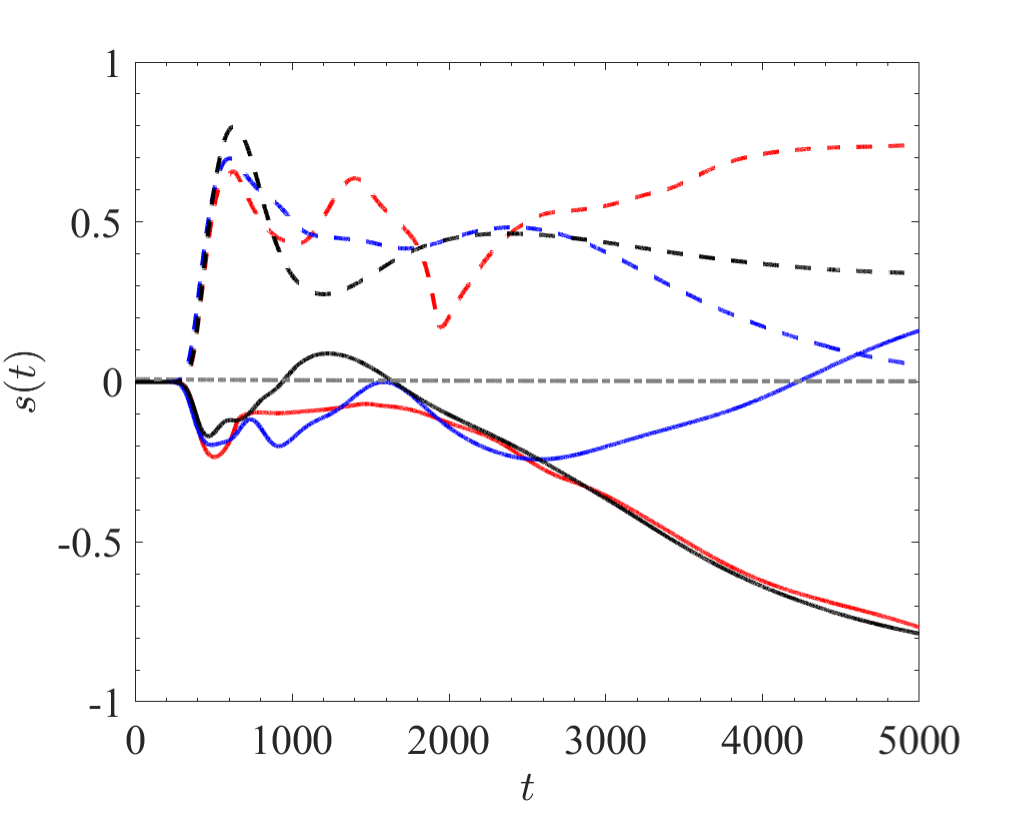}
	\caption{Temporal evolution of skewness of more ($R = 3$, solid lines) and less ($R = -3$, dashed lines) viscous samples for different flow conditions: Type-I (blue), Type-II (black), and Type-III (red). The grey dash-dotted line represents the stable scenario ($R = 0$).  Other parameters are the same as in the figure \ref{Fig:Type-I}.} 
	\label{Fig:skewness}
\end{figure}

Central moments of the transverse-averaged concentration profile help better understand the spreading and mixing of the sample in the longitudinal directions. The first and second central moments are already defined in equations \eqref{eq:first_moment_x} and \eqref{eq:var_x}. The third central moment, defined as
 \begin{equation}
     \label{eq:third_moment}
     s(t)=\frac{\int_{0}^{L_x}(x-m_x(t))^3\bar{c}(x,t){\rm d} x/\int_{0}^{L_x} \bar{c}(x,t) {\rm d} x}{\sigma_x^3},
     \end{equation}
captures the asymmetry of the solute about its mean position. Temporal evolutions of the deviation of the mean solute concentration from its initial position (see figure \ref{Fig:mean}(a)) and the spreading of the solute (see figure \ref{Fig:mean}(b)) depict that a less viscous samples traverses and spreads more as compared to its more viscous counterpart under all the three flow conditions considered here. This is because viscous fingers carry the less viscous sample along with the flow, whereas the more viscous sample is carried against the flow. The addition of solute mass in the system across the transverse boundaries redistributes the solute along the flow. Since diffusion spreads the sample symmetrically about its centre, skewness remains zero throughout. Figure \ref{Fig:skewness} depicts that a less viscous sample is right skewed for all the flow conditions; whereas, a more viscous sample is mostly left skewed. In summary, transverse periodicity makes the sample less skewed compared to the other two flow conditions. 

\section{Conclusion} \label{sec:slice_conclusion}

This study provides a comprehensive analysis of the effects of viscosity contrast and boundary conditions on miscible viscous fingering of a finite slice in homogeneous isotropic porous media. We utilized a fourth-order accurate compact finite difference method and Crank-Nicolson time stepping to solve the coupled nonlinear partial differential equations of advection-diffusion type. It is observed that irrespective of the choice of the boundary conditions, the finite slice undergoes fingering deformation at the rear interface for $R > 0$, whereas viscous fingering deforms the frontal interface for $ R < 0$. In the latter, viscous fingers lead to faster and more extensive growth of the instability due to its movement in the flow direction, while for $R > 0$, fingering at the rear interface results in slower growth. Although the onset of viscous fingering is unaffected by whether the transverse boundaries are permeable or impermeable, the late-time dynamics--particularly the fingering behavior and solute transport after breakthrough--depend on the imposed boundary conditions. Permeable boundaries (Type-III boundary conditions) allows mass enhancement and subsequently promotes larger mixing lengths compared to other types of transverse boundaries: periodic (Type-I) and impermeable (Type-II). The non-monotonic interfacial behaviour observed under the Type-III condition further distinguishes this case, highlighting the complex role of boundary effects in shaping the flow dynamics. These findings have important implications for applications in chromatographic separation and pollution spread, where control of VF can be crucial to the efficiency of fluid displacement processes.

% Use the style of numbering in square brackets.
% If nothing is used, default style will be taken.
%\begin{enumerate}[a)]
%\item 
%\item 
%\item 
%\end{enumerate}  

%\begin{itemize}
%\item 
%\item 
%\item 
%\end{itemize}  

%\begin{description}
%\item[]
%\item[] 
%\item[] 
%\end{description}  

\section*{Acknowledgments}

Authors acknowledge financial support through the Core Research Grant (CRG/2023/004156) supported by the Science and Engineering Research Board, Department of Science and Technology, Government of India. 
S.P. acknowledges financial support through the Start-Up Research Grant (SRG/2021/001269), MATRICS Grant (MTR/2022/000493) from the Science and Engineering Research Board, Department of Science and Technology, Government of India, and Start-up Research Grant (MATHSUGIITG01371SATP002), IIT Guwahati. 

% \clearpage %%Remove this from your manuscript

% Figure
% \begin{figure}%[]
%   \centering
% %    \includegraphics{}
%     \caption{}\label{fig1}
% \end{figure}

% \begin{table}%[]
% \begin{tabular*}{\tblwidth}{@{}LL@{}}
% \toprule
% \midrule
%  & \\
%  & \\
%  & \\
%  & \\
% \bottomrule
% \end{tabular*}
% \end{table}

% Uncomment and use as the case may be
%\begin{theorem} 
%\end{theorem}

% Uncomment and use as the case may be
%\begin{lemma} 
%\end{lemma}


% \section{}\label{}

\printcredits

%\bibliographystyle{cas-model2-names}
%
%% Loading bibliography database
%\bibliography{mybib}



\end{document}